\documentclass[conference,letterpaper]{IEEEtran}
\usepackage{preprint}

\newcommand{\aflow}{AFlow}
\newcommand{\enigmata}{Enigmata}
\newcommand{\autoresearch}{AutoResearch}

\newcommand{\ba}{\boldsymbol{a}}
\newcommand{\bt}{\boldsymbol{t}}
\newcommand{\bm}{\boldsymbol{m}}
\newcommand{\temp}{$t$}
\newcommand{\cmark}{\textcolor{green!50!black}{\ding{51}}}
\newcommand{\xmark}{\textcolor{red!65!black}{\ding{55}}}

\newcommand{\symPuz}{p}
\newcommand{\symBun}{b}
\newcommand{\symOpt}{o}
\newcommand{\symWrk}{w}
\newcommand{\symAgg}{g}
\newcommand{\symSeed}{K}
\newcommand{\symMcts}{N}

\newcommand{\valPuz}{30}

\newcommand{\valOpt}{1}
\newcommand{\valWrk}{2}
\newcommand{\valAgg}{2}
\newcommand{\valSeed}{5}
\newcommand{\valMcts}{20}
\newcommand{\valInst}{20}

\title{Mining the Design Space\\of Multi-Agent Workflows}
\author{%
  \IEEEauthorblockN{%
  Xavier Cadet\textsuperscript{\affiliation{\alpha},*}\quad
  Junyan Cheng\textsuperscript{\affiliation{\alpha},*}\quad
  Peter Chin\textsuperscript{\affiliation{\alpha}}\quad
  Kyle Richardson\textsuperscript{\affiliation{\beta}}}
  \IEEEauthorblockA{%
  \textcolor{accent}{\textsuperscript{\ensuremath{\alpha}}Dartmouth College}\qquad
  \textcolor{accent}{\textsuperscript{\ensuremath{\beta}}Allen Institute for AI}\\[2pt]
  {\small\textsuperscript{*}Equal contribution.}\\[3pt]
  {\small\texttt{\{xavier.fjf.cadet,jc.th\}@dartmouth.edu}\quad
  \texttt{kyler@allenai.org}}\\[4pt]
  \href{https://junyan.ch/apex-paper/}{\textbf{Website}\enspace junyan.ch/apex-paper}
  \qquad
  \href{https://github.com/xcadet/apex}{\textbf{Code}\enspace github.com/xcadet/apex}}%
}
\date{}
\hypersetup{
  pdftitle={Mining the Design Space of Multi-Agent Workflows},
  pdfauthor={Xavier Cadet, Junyan Cheng, Peter Chin, Kyle Richardson},
  pdfsubject={Mining reproducible design principles from multi-agent workflow search traces},
  pdfkeywords={large language models, multi-agent systems, workflow optimization, design-space mining}
}

\begin{document}
\maketitle
\thispagestyle{plain}
\begingroup
\renewcommand{\thefootnote}{\fnsymbol{footnote}}
\footnotetext[1]{Equal contribution. XC: system implementation \& operator set,
large-scale experiments, data analysis, and empirical write-up;
JC: project conception, theoretical framework \& instruments,
experimental design, and theoretical write-up.}
\endgroup

\begin{abstract}
Designing large language model (LLM) multi-agent systems (MAS) still relies on trial and error and informal ``folk wisdom.'' Manual studies test only a few communication structures or topologies (e.g., sequential, parallel, shared-memory, or debate), while automated optimizers discard search trajectories after selecting a task-specific workflow. We present a two-phase pipeline for mining reproducible design principles from workflow code and execution traces. We repurpose AFlow's MCTS optimizer across six composable operators and 26 deterministically verified Enigmata tasks, retaining all intermediate rounds to build a corpus of 28,568 executable workflows. A sandboxed AutoResearch loop then mines orchestration code and prompt templates under an $L_0$-penalized validation objective and tests candidate principles on held-out searches. Across 300 feature-mining trials, we identify effects that persist on held-out searches across three levels of workflow design: workflows beginning with independent solvers and one-shot aggregation (\textsc{Star}-first) perform worse ($\Delta=-0.045$); persistent shared documents (\textsc{SharedMemory}) outperform debate ($\Delta=+0.135$); and prompt specificity and placeholder-based context routing yield gains comparable to communication-structure selection ($\Delta=+0.118$). More broadly, our pipeline and results show the feasibility of turning discarded search trajectories into a statistical testbed for discovering and validating multi-agent workflow design principles.

\end{abstract}

\section{Introduction}

Large Language Model (LLM)–based Multi-Agent Systems (MAS) have become a foundational tool for complex problem solving in many areas such as software engineering \citep{yang2024swe}, economic forecasting \citep{analytica,aapm}, and scientific discovery \citep{lu2026towards,novikov2025alphaevolve}.
In these systems, performance depends not only on the base LLM, but critically on the workflow architecture: how agents are connected, how intermediate state is routed, and how role-specific prompts are structured \citep{scalingagent}.
Yet, designing multi-agent workflows remains largely guided by manual trial-and-error and informal ``folk wisdom'' (e.g., ``more agents improve accuracy'' or ``multi-agent debate improves reasoning'').
This raises the question: \textit{Can workflow search traces be used to identify statistically reliable principles for multi-agent system design?}

\begin{figure*}[t]
    \centering
    \includegraphics[width=\linewidth]{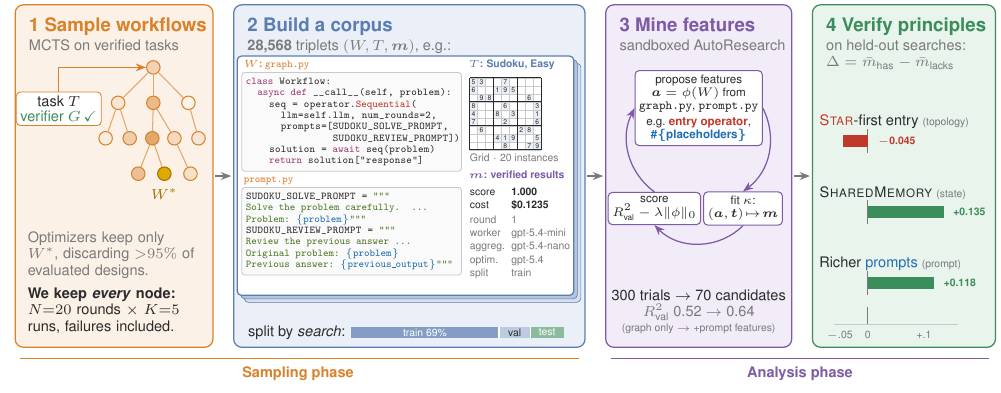}
    \caption{
    \textbf{Overview of our two-phase pipeline.}
    \textbf{(1)~Sample:} for each verified task bundle $T$, an \aflow{} MCTS search proposes and evaluates workflows; optimizers keep only the final $W^{*}$, whereas we keep \emph{every} evaluated node ($N{=}20$ rounds $\times$ $K{=}5$ runs, failures included).
    \textbf{(2)~Build a corpus:} each of the $28{,}568$ entries is a triplet $(W, T, \bm)$, an executable workflow $W$ (orchestration code \texttt{graph.py} plus its prompt templates \texttt{prompt.py}), a task bundle $T$, and verified metrics $\bm$ (score, cost), split into train/val/test at the level of whole searches.
    \textbf{(3)~Mine features:} a sandboxed \autoresearch{} loop proposes code and prompt features $\ba=\phi(W)$ (e.g., the entry operator or the number of \texttt{\{placeholders\}}), fits a regressor $\kappa$ from $(\ba,\bt)$ to $\bm$, and scores candidates by $L_0$-penalized validation $R^2$.
    \textbf{(4)~Verify principles:} surviving features are tested on held-out searches as score contrasts $\Delta$ between workflows that have vs.\ lack the feature.
    }
    \label{fig:overview}
\end{figure*}

Recent studies have taken initial steps toward quantifying MAS design choices, examining agent-count scaling \citep{zhang2024more}, load balancing in compressor-predictor pipelines \citep{he2025information}, and canonical topologies \citep{scalingagent}.
However, two bottlenecks limit existing paradigms:
(1)~\emph{manual empirical studies} evaluate only a handful of fixed, hand-crafted topologies (typically 3--5 patterns), leaving composed workflows and sub-topological prompt implementations unexamined; and
(2)~\emph{automated workflow optimizers} such as \aflow{} \citep{aflow} and ADAS \citep{hu2025automated} dynamically search over code and prompts, but discard over $95\%$ of evaluated architectures to return a single task-specific optimum $W^*$.

To bridge this gap, we introduce a two-phase empirical pipeline that transforms complete workflow-search trajectories into a controlled testbed for discovering design principles (Fig.~\ref{fig:overview}). In the \textbf{Sampling Phase}, we repurpose the MCTS-based \aflow{} optimizer \citep{aflow} to compose six workflow operators: \textsc{Sequential}, \textsc{Star}, \textsc{SharedMemory}, \textsc{Debate}, \textsc{StarWithCoordinator}, and \textsc{DebateWithModerator}. Rather than retaining only the final workflow, we collect all $N=20$ evaluated workflows from each of $K=5$ independent runs per condition. To limit confounding from external tools, specialized domain knowledge, and LLM-as-a-judge noise, we evaluate every workflow in a tool-free setting on 26 deterministically verified reasoning tasks from \enigmata{} \citep{enigmata}. This produces a corpus of $28{,}568$ executable workflow--task--performance triplets.

In the \textbf{Analysis Phase}, we run an adapted \autoresearch{} loop \citep{autoresearch} in a leakage-controlled sandbox with a strict search-level train/validation/test split. Across 300 trials, it mines features from orchestration code (\texttt{graph.py}) and role-prompt templates (\texttt{prompt.py}) under an $L_0$-regularized validation $R^2$ objective; we then test 70 candidate contrasts on held-out searches.

Our analysis identifies three effects that persist on held-out searches, spanning progressively finer levels of workflow specification: (1)~\emph{Entry-Point Topology:} workflows beginning with \textsc{Star} fan-out---in which parallel workers solve independently before a one-shot aggregator combines their outputs---perform worse than alternative entry structures (median $\Delta=-0.045$), corroborating \citet{scalingagent}'s finding that uncoordinated agents can amplify errors; (2)~\emph{Communication State:} \textsc{SharedMemory}, in which workers sequentially update a persistent shared document \citep{hayes1985blackboard}, outperforms conversational message passing ($\Delta=+0.135$); and
(3)~\emph{Prompt Structure:} more specific prompt templates and named placeholders that route context lift validation $R^2$ from $0.521$ with graph features alone to $0.638$, with held-out gains ($\Delta=+0.118$) comparable to choosing the overall communication or orchestration topology.

Taken together, our work makes the follow three concrete contributions\footnote{Our full code and dataset will be publicly released upon publication, along with all discovery artifacts.}:
\begin{enumerate}\setlength{\itemsep}{1pt}\setlength{\parskip}{0pt}\setlength{\topsep}{2pt}
\item \textbf{Full-Trace Design-Space Mining.} We introduce a pipeline that combines complete MCTS search trajectories with deterministic evaluation to mine code- and prompt-level design principles (\S~\ref{sec:framework}, \S~\ref{sec:method}).

\item \textbf{Large-Scale Workflow Corpus.} We construct a corpus of 28,568 executable multi-agent workflows covering 26 verifiable reasoning tasks, with train/validation/test splits defined at the level of complete searches (\S~\ref{subsec:apexdb}).

\item \textbf{Multi-Granularity Design Evidence.} We identify and validate held-out effects across three levels of workflow design: entry-point topology, communication state, and prompt structure (\S~\ref{subsec:analysis}).
\end{enumerate}

\section{Related Work}

\paragraph{Multi-Agent System Design and Topologies}
Designing effective multi-agent LLM workflows has largely relied on practitioner heuristics \citep{wiesinger2024agents,chenAgentVerseFacilitatingMultiAgent2023}. 
While classical AI established blackboard architectures \citep{hayes1985blackboard} where specialist modules read and write to a shared global state, recent LLM-MAS studies have begun quantifying specific design dimensions: \citet{zhang2024more} and \citet{qianSCALINGLARGELANGUAGE2025} studied agent-count and collaboration scaling; \citet{he2025information} analyzed load balancing in compressor-predictor pipelines; \citet{chen2026diversity} examined diversity collapse; \citet{willis2026evaluating} studied large agent swarms; and \citet{scalingagent} benchmarked five canonical MAS patterns. 
However, these works evaluate a small, fixed set of hand-crafted topologies and treat prompts as fixed constants. 
By contrast, we mine empirical regularities across tens of thousands of dynamically composed workflows, jointly analyzing macroscopic topologies and sub-topological prompt structure.

\begin{figure*}[t]
    \centering
    \includegraphics[width=\linewidth]{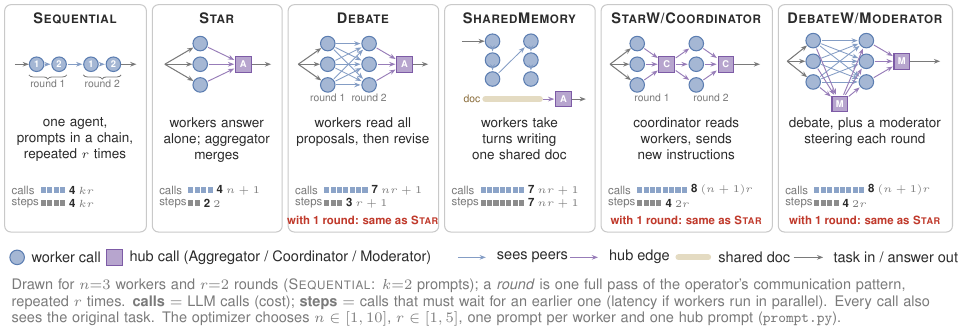}
    \caption{
    \textbf{The six operator primitives $\mathcal{O}$ spanning our design space}, drawn for $n{=}3$ workers and $r{=}2$ rounds.
    Circles are worker calls and squares are hub calls (aggregator~A, coordinator~C, moderator~M); blue edges show which earlier outputs a call reads, purple edges connect workers to a hub.
    \textsc{Sequential} chains prompts through one agent; \textsc{Star} fans out to isolated workers and aggregates once; \textsc{Debate} lets every worker read all peer proposals before revising; \textsc{SharedMemory} has workers take turns appending to one persistent document (blackboard); \textsc{StarW/Coordinator} and \textsc{DebateW/Moderator} add a hub call after every round.
    \emph{calls} counts LLM calls (cost) and \emph{steps} counts sequentially dependent calls (latency); with a single round the three multi-round operators reduce to \textsc{Star}.
    The optimizer composes operators in \texttt{graph.py} and writes their prompts in \texttt{prompt.py} (Fig.~\ref{fig:composition}).
    }
    \label{fig:operators}
\end{figure*}

\paragraph{Automated Workflow and Prompt Search}
Automated search methods have targeted prompt optimization \citep{khattab2023dspy,guo2024connecting,gepa}, graph connections \citep{zhuge2024gptswarm,zhang2025gdesigner}, and agent code evolution \citep{zhang2026darwin}. 
Full-workflow optimizers, including MCTS-based \aflow{} \citep{aflow}, MASS \citep{mass}, ADAS \citep{hu2025automated}, and supernet routing \citep{supernet}, iteratively mutate workflow graphs and prompts to maximize accuracy. 
Whereas these methods discard intermediate candidates to output a single workflow $W^*$, we repurpose MCTS workflow search to harvest complete search trajectories for population-level analysis.

\paragraph{Controlled Reasoning Evaluation}
Downstream agentic benchmarks in software engineering \citep{jimenez2023swe,yang2024swe}, API/web interaction \citep{trivedi2024appworld,apeiron}, and scientific research \citep{bragg2026astabench} conflate multi-agent coordination with domain knowledge, external tool reliability, and LLM-as-a-judge variance \citep{scalingagent}. 
To isolate reasoning and coordination, synthetic benchmarks use verifiable problems in constraint satisfaction \citep{zebralogic}, combinatorics \citep{mittal2024fcorebench}, and SAT \citep{wei2025satbench}. 
We evaluate workflows on \enigmata{} \citep{enigmata}, leveraging its parametric generators and deterministic verifiers across difficulty tiers.

\section{Formulating Design-Space Mining}
\label{sec:framework}

Given a collection of executable multi-agent workflows, each evaluated on a concrete reasoning task $T$ (e.g., a bundle of Sudoku instances checked by a verifier), we ask: \emph{what structural properties of the workflow designs and prompts reliably predict higher task performance?} In this section, we introduce our workflow representations and mining objective (\S~\ref{subsec:formulate}), as well as outline the three  operational challenges motivating our pipeline in \S~\ref{subsec:bias}.

\subsection{Workflow Representation and Mining Objective}
\label{subsec:formulate}

Following \citet{scalingagent} and \citet{aflow}, we represent an evaluated agentic system as $S = (W, E)$, pairing an executable \textbf{workflow} $W \in \mathcal{W}$ with an \textbf{evaluation environment} $E = (T, G)$. We consider each part in turn.

\paragraph{Executable Workflows ($W$) and Environment ($E$)}
A workflow $W = (\mathcal{N}, C, \Omega)$ is operationalized as an executable Python program comprising two coupled source files: an orchestration graph (\texttt{graph.py}) and a prompt module (\texttt{prompt.py}).
Here, $\mathcal{N} = \{\nu_1, \nu_2, \dots\}$ denotes the LLM-invoking agent calls, each parameterized by a backbone model $M_i$, a text prompt template $P_i$ (with runtime placeholders such as \texttt{\{problem\}} or \texttt{\{worker\_outputs\}}), and temperature $\tau_i$.
The communication topology $C$ and message-routing policy $\Omega$ are expressed by composing multi-agent \emph{operator primitives} from a vocabulary $\mathcal{O}$ (Fig.~\ref{fig:operators}; parameters in Table~\ref{tab:operators}, \S~\ref{apdx:operator_set}).
The environment $E = (T, G)$ pairs a \emph{task bundle} $T$ (i.e., \emph{$M$ problem instances from a given reasoning task and difficulty tier}) with a deterministic verifier $G$.
Executing $W$ on $T$ yields a performance vector $\bm = G(W, T) \in \mathcal{M}$, recording mean verification accuracy ($\text{score} \in [0, 1]$) and inference cost (USD).

\paragraph{Design-Space Feature Mining}
In the \emph{sampling phase}, we collect evaluated workflow triplets $\mathcal{D} = \{(W_i, T_i, \bm_i)\}_{i=1}^{|\mathcal{D}|}$.
Because each $W_i$ is raw source code, in the \emph{analysis phase} we search for a programmatic feature extractor $\phi: \mathcal{W} \to \mathbb{R}^d$ that parses \texttt{graph.py} and \texttt{prompt.py} into an interpretable feature vector $\ba = \phi(W)$ (e.g., operator counts, entry-point topology, or prompt template length and placeholder count), alongside task descriptors $\bt = \psi(T)$ and a regressor $\kappa$.
Optimizing $(\phi, \psi, \kappa) \in \mathbf{H}$ on train/validation splits yields candidate features $\mathcal{H} \subset \mathbf{H}$.
Each candidate is then verified on held-out test searches by splitting the workflows of every puzzle into two \emph{arms}, as in an A/B test, a \emph{high arm} in which the feature is present (or above the per-puzzle median) and a \emph{low arm} in which it is absent (or at/below the median), and estimating the per-puzzle contrast $\Delta = \mathbb{E}[\text{score} \mid \text{high arm}] - \mathbb{E}[\text{score} \mid \text{low arm}]$ (\S~\ref{subsec:analysis}).

\subsection{Methodological Challenges in Design-Space Mining}
\label{subsec:bias}

Extracting reliable empirical principles from observational workflow data requires overcoming three operational challenges (summarized in Table~\ref{tab:bias-component-map} and formalized in Appendix~\ref{app:theory}):

\paragraph{Challenge 1: Design-Space Coverage vs.\ Implementation Competence}
Mining principles requires a sample that both \emph{covers} many distinct designs and consists of \emph{competent} implementations.
The naive route to coverage, generating workflow programs from scratch (e.g., prompting an LLM to write arbitrary orchestration code and prompts without constraints), fails the second requirement: most candidates break with syntax errors or degenerate prompts, burying topological effects under implementation noise.
Conversely, keeping only an optimizer's final argmax workflow $W^*$ discards the negative and intermediate examples needed to estimate contrasts.
We therefore repurpose the MCTS-based \aflow{} optimizer (\S~\ref{subsec:traverser}) over a structured 6-operator vocabulary $\mathcal{O}$ and retain \emph{all} $N=20$ intermediate workflows per search under a consistent LLM optimizer policy.

\paragraph{Challenge 2: Domain, Tool, and Grading Confounds}
On downstream benchmarks such as software engineering or web navigation, accuracy is conflated with domain knowledge, external tool reliability, and LLM-as-a-judge grading noise \citep{scalingagent}.
To isolate multi-agent coordination and prompt structure, we evaluate workflows in a tool-free setting using \enigmata{} (\S~\ref{subsec:evaluation}), where tasks are self-contained abstract reasoning puzzles scored by exact, deterministic rule-based verifiers on fixed instance bundles.

\paragraph{Challenge 3: High-Dimensional Feature Overfitting and Leakage}
Because infinitely many features can be parsed from \texttt{graph.py} and \texttt{prompt.py}, manual feature selection is narrow, whereas unconstrained search can overfit or exploit dataset artifacts (such as MCTS round indices or within-search correlation).
We address this via an $L_0$-regularized validation objective (\S~\ref{subsec:autoresearch}) inside a strict sandbox, partitioning $\mathcal{D}$ into train, validation, and test splits at the \emph{\aflow{} search level} and stripping all search/round identifiers from workflow files.

\section{Methodology: Sampling and Analysis Phases}
\label{sec:method}

\begin{figure}[t]
  \centering
  \includegraphics[width=\columnwidth]{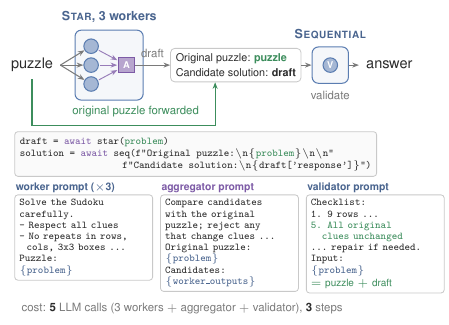}
  \caption{\textbf{Composing operators in \texttt{graph.py}.} A \textsc{Star} workflow (three workers, one aggregator) drafts a Sudoku solution; \texttt{graph.py} then forwards the \emph{original} puzzle together with the draft into a one-step \textsc{Sequential} validator. Every call's prompt template lives in \texttt{prompt.py}, with placeholders such as \texttt{\{problem\}} and \texttt{\{worker\_outputs\}} filled at run time. Features such as the entry operator, chain length, and placeholder counts are parsed from exactly these two files.}
  \label{fig:composition}
\end{figure}

\begin{figure*}[t]
  \centering
  \begin{minipage}[t]{0.48\textwidth}
    \centering
    \includegraphics[width=\linewidth]{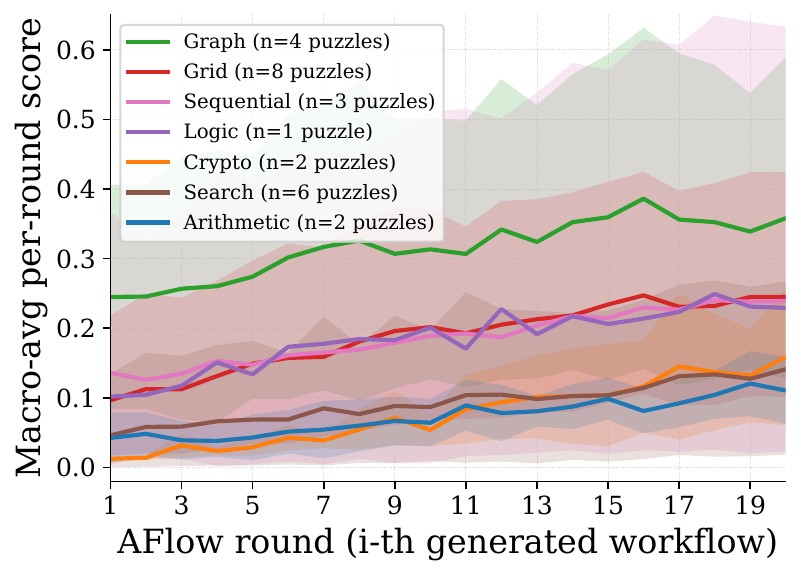}
    \caption{\textbf{Full-Trace \aflow{} Search.} Macro-mean raw score per MCTS round by puzzle category ($1{,}701$ searches, $26$ puzzles; $95\%$ bootstrap CIs); retaining all 20 rounds captures a wide performance trajectory.}
    \label{fig:first}
  \end{minipage}
  \hfill %
  \begin{minipage}[t]{0.48\textwidth}
    \centering
    \includegraphics[width=\linewidth]{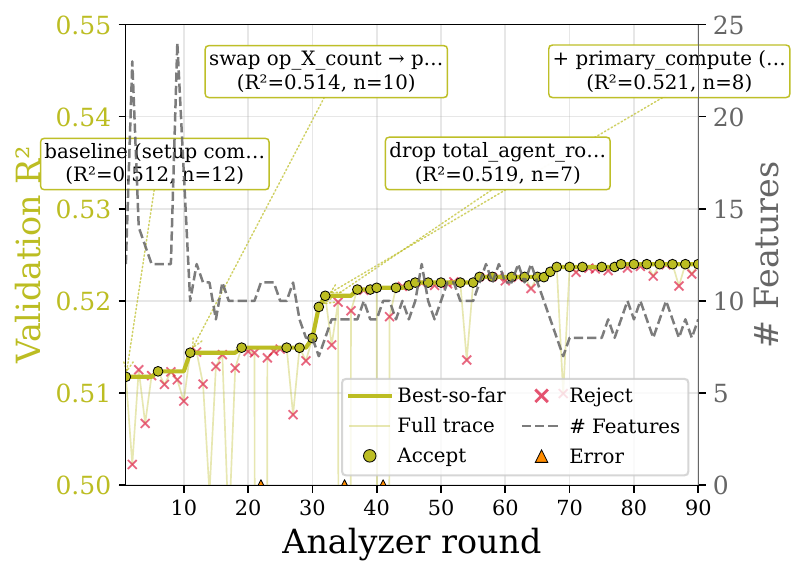}
    \caption{\textbf{\autoresearch{} Feature-Mining Trace.} Validation $R^2$ (left axis) and active feature count $\|\phi\|_0$ (right axis) over iterations; the $L_0$-penalized objective improves fit while pruning redundant features.}
    \label{fig:second}
  \end{minipage}
\end{figure*}

Our pipeline operates in two stages: a \textbf{sampling phase} that pairs full-trace \aflow{} workflow search (\S~\ref{subsec:traverser}) with deterministic \enigmata{} evaluation (\S~\ref{subsec:evaluation}), followed by an \textbf{analysis phase} that mines workflow code and prompt features via a sandboxed \autoresearch{} loop (\S~\ref{subsec:autoresearch}).

\subsection{Sampling Phase: Full-Trace Workflow Search via \aflow{}}
\label{subsec:traverser}

In \aflow{} \citep{aflow}, each workflow $W$ is represented as an executable Python class in \texttt{graph.py} accompanied by role-specific hard text prompt templates in \texttt{prompt.py} (no soft prompts or gradient updates). 
Given a task bundle $T$, \aflow{} searches over workflows across $N = 20$ MCTS rounds:
(i)~\emph{Selection:} A parent workflow is sampled from the top-$k$ ($k=4$) evaluated workflows using a mixed uniform/softmax distribution ($\lambda = 0.2$ uniform weight, temperature $\alpha = 0.4$; Eq.~\ref{eq:mixture} in \S~\ref{app:sampling}) to prevent premature collapse;
(ii)~\emph{Expansion:} An LLM optimizer (fixed to \texttt{GPT-5.4} at temperature $0$) inspects the parent's code and score history and proposes an edit that adds, removes, or rechains operators in \texttt{graph.py} and refines role prompts in \texttt{prompt.py} (\S~\ref{app:prompts});
(iii)~\emph{Evaluation:} The child workflow is executed on the $M=20$ instances in $T$; and
(iv)~\emph{Backpropagation:} Scores and costs are recorded in the search tree.
Steps (i)--(iv) are the unmodified \aflow{} loop; we change three things (\S~\ref{apdx:aflow}).
First, the operator registry is replaced by our six primitives (Fig.~\ref{fig:operators}), so every proposed edit is a composition of known topologies.
Second, evaluation targets fixed \enigmata{} bundles with deterministic verifiers (\S~\ref{subsec:evaluation}).
Third, unlike point optimization that returns only $W^*$, we retain \emph{all} $N = 20$ workflows per trace across $K = 5$ independent runs per condition, capturing both successful and failed designs below the best-so-far envelope (Fig.~\ref{fig:first}; single tree in Fig.~\ref{fig:single_aflow_trace}, \S~\ref{apdx:traces}).

\paragraph{Composable Operator Primitives and Roles}
To span established coordination patterns while permitting multi-stage compositions, we equip the search space with six operator primitives $\mathcal{O}$ (Fig.~\ref{fig:operators}, Table~\ref{tab:operators}; \S~\ref{apdx:operator_set}), synthesized from recent MAS surveys \citep{scalingagent,mass,supernet,chenAgentVerseFacilitatingMultiAgent2023}: single-agent linear chaining (\textsc{Sequential}), isolated parallel fan-out followed by one-shot aggregation (\textsc{Star}), blackboard-style sequential writing to a persistent shared document (\textsc{SharedMemory}; \citealp{hayes1985blackboard}), all-to-all peer proposal exchange (\textsc{Debate}), and hub-and-spoke or moderated multi-round variants (\textsc{StarWithCoordinator}, \textsc{DebateWithModerator}). 
An \emph{operation} in \texttt{graph.py} is an asynchronous call to one operator primitive with optimizer-chosen parameters: $n \in [1, 10]$ workers, $r \in [1, 5]$ \emph{rounds}, and custom prompts in \texttt{prompt.py}.
A round is one complete pass of the operator's communication pattern, e.g., every \textsc{SharedMemory} worker reads the document once and appends to it, or every \textsc{Debate} worker reads its peers' proposals once and revises, and an operator with $r > 1$ repeats this pass $r$ times, each pass conditioned on the previous one (Fig.~\ref{fig:operators} is drawn for $r{=}2$); these operator rounds are distinct from the $N{=}20$ MCTS \emph{search} rounds above.
Operations can be chained sequentially or conditionally, with the outputs of one operator (and the original problem) forwarded into the prompts of the next (e.g., $\textsc{Star} \to \textsc{Sequential}$ in Fig.~\ref{fig:composition}; the most common chains are listed in Table~\ref{tab:most_common_operator_combinations}). 
Each operator uses up to two LLM roles (a \emph{worker} and an \emph{aggregator/coordinator/moderator}) drawn from $\{\texttt{GPT-5.4-mini}, \texttt{GPT-5.4-nano}\}$, with at most $20$ LLM calls per workflow execution (\S~\ref{apdx:exp}).

\subsection{Sampling Phase: Controlled Reasoning Evaluation via \enigmata{}}
\label{subsec:evaluation}

All agents operate in a \emph{pure-reasoning, tool-free setting} (no code interpreters or search tools), so collaboration acts only through task decomposition, diverse candidate generation, peer verification, and shared state tracking. 
We select all $30$ tasks in \enigmata{} \citep{enigmata} (\emph{Crypto, Arithmetic, Logic, Grid, Graph, Search, and Sequential} categories) with programmatic generators and deterministic rule-based verifiers (exact match $\in \{0, 1\}$), yielding $87$ (puzzle, difficulty) bundles across \{Easy, Medium, Hard\}. 
For each bundle $T$, we deterministically generate $M = 20$ instances and reuse this fixed set across all $K = 5$ searches and $N = 20$ rounds within that bundle so score differences reflect workflow structure rather than instance sampling noise (\S~\ref{app:enigmata}). 
Filtering $1{,}031$ syntax-invalid workflows ($3.0\%$) and excluding $4$ sliding-tile puzzles where all single-agent baselines and workflows scored $0.00$ (\S~\ref{app:failures}) yields $28{,}568$ executable workflows across $26$ puzzles.

\subsection{Analysis Phase: Sandboxed Feature Mining via \autoresearch{}}
\label{subsec:autoresearch}

To extract design principles from \texttt{graph.py} and \texttt{prompt.py}, we adapt \autoresearch{} \citep{autoresearch} as an automated feature-mining loop. 
At each iteration, an LLM agent edits a feature-extraction and regression script (\texttt{train.py}) that parses workflow source files into features $\phi(W) \in \mathbb{R}^d$, joins them with task descriptors $\psi(T)$, and fits per-puzzle regressors $\kappa$ on $D_{\text{train}}$. 
To enforce parsimony and mitigate overfitting, each candidate is scored on the held-out validation split $D_{\text{val}}$ via an $L_0$-regularized \textbf{validation objective} $\mathcal{F}(\phi, \psi, \kappa)$ defined as:
\begin{equation}
\label{eq:srm}
\begin{aligned}
\mathcal{F}(\phi,\psi,\kappa)
  &= R^2_{D_{\mathrm{val}}}(\bm,\widehat{\bm}) - \lambda\|\phi\|_0, \\
\widehat{\bm} &= \kappa\bigl(\phi(W),\psi(T)\bigr).
\end{aligned}
\end{equation}
where $\|\phi\|_0 = d$ penalizes active feature count (Fig.~\ref{fig:second}; multi-run comparison in Fig.~\ref{fig:autoresearch_trace}).

\paragraph{Search-Level Splitting and Sandbox Isolation}
Because workflows within the same MCTS trace are evolutionary descendants, we partition the corpus into training ($69.4\%$), validation ($15.0\%$), and held-out test ($15.6\%$) splits strictly at the \emph{\aflow{} search level} (Table~\ref{tab:dataset_structure}; \S~\ref{apdx:split}). 
The feature-mining loop operates in a read-only sandbox that strips all search hashes, MCTS round indices, and file paths from workflow sources, forbids target-derived encodings, and physically isolates $D_{\text{test}}$ for final verification (\S~\ref{app:caveats}).

\section{Empirical Analysis}
\label{sec:analysis}

\begin{figure*}[t]
    \centering
    \begin{minipage}{0.96\linewidth}
        \centering
        {\scriptsize\sffamily\bfseries
        \hspace*{0.04\linewidth}\makebox[0.29\linewidth][c]{Principle 1: Star-first hurts}\hfill
        \hspace*{0.02\linewidth}\makebox[0.30\linewidth][c]{Principle 2: SharedMemory helps}\hfill
        \hspace*{0.02\linewidth}\makebox[0.29\linewidth][c]{Principle 3: Richer prompt helps}}\\[2pt]
        \includegraphics[trim=0 0 0 24.5bp, clip, width=\linewidth]{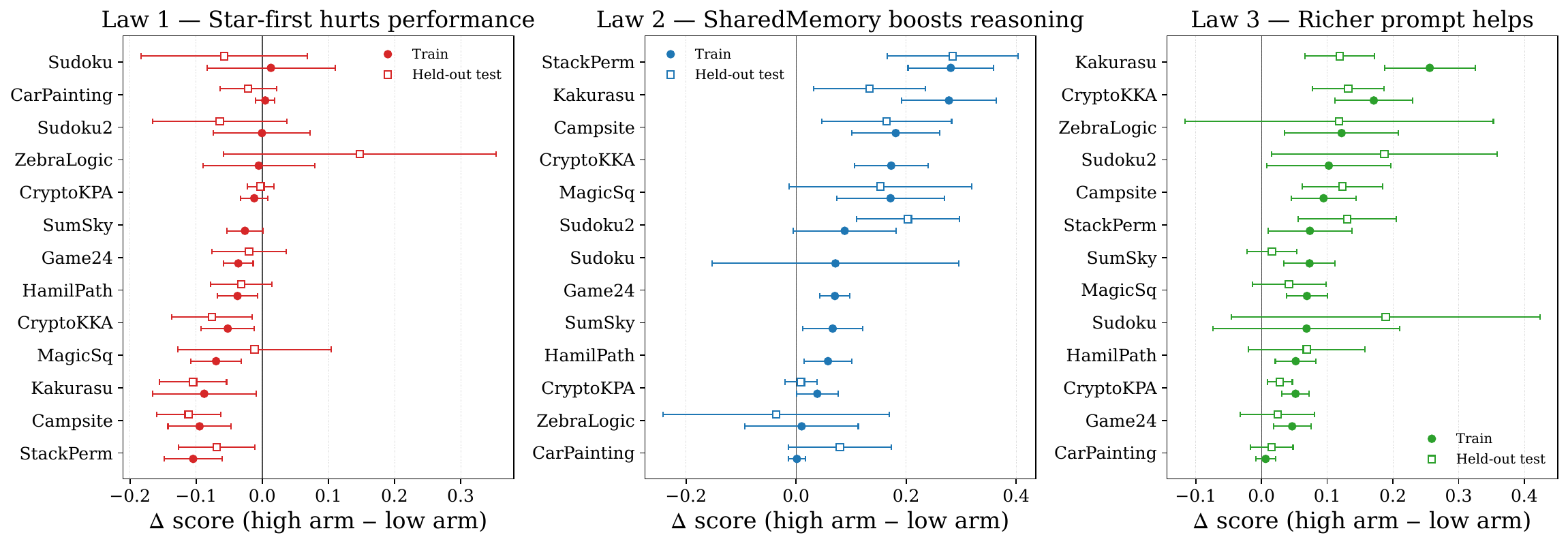}
    \end{minipage}
    \caption{\textbf{Per-Puzzle Effect Sizes of Mined Design Principles.} Per-puzzle score contrast $\Delta$ (high arm $-$ low arm) for the three principles on the search-level train split (filled circles, $n = 19{,}822$) and held-out test split (open squares, $n = 4{,}468$), with $95\%$ cluster-robust CIs; effect directions are consistent out-of-sample.}
    \label{fig:effect}
\end{figure*}

\begin{table*}[t]
\centering
\footnotesize
\setlength{\tabcolsep}{4.5pt}
\renewcommand{\arraystretch}{1.0}
\begin{tabular}{@{}l l c c c c c c c c c c c c c c@{}}
\toprule
 &  & \multicolumn{5}{c}{Train} & \multicolumn{5}{c}{Held-out test} & \multicolumn{2}{c}{Sign agreement} & \multicolumn{2}{c}{Test-verified} \\
\cmidrule(lr){3-7} \cmidrule(lr){8-12} \cmidrule(lr){13-14} \cmidrule(lr){15-16}
 & Variant & $n$ & $k_+$ & $k_-$ & med $\Delta$ & $p_{\mathrm{W}}$ & $n$ & $k_+$ & $k_-$ & med $\Delta$ & $p_{\mathrm{W}}$ & $k/n_{\cap}$ & $p_{\mathrm{B}}^{\cap}$ & $k/n_{\mathrm{tr}}$ & $p_{\mathrm{B}}^{\mathrm{tr}}$ \\
\midrule
\multicolumn{16}{l}{\textbf{\textit{Principle 1 (RQ1, Entry-Point Topology): Star-first hurts performance}} \textit{(claim direction $-$)}} \\
 & Star 1st & 13 & 0 & 7 & $-0.036$ & .008 & 12 & 0 & 4 & $-0.045$ & .038 & 9/12 & $<$.001 & 9/10 & .011 \\
 & Star$\times$compute & 13 & 0 & 7 & $-0.036$ & .008 & 12 & 0 & 3 & $-0.041$ & .038 & 9/12 & $<$.001 & 9/10 & .011 \\
\addlinespace[2pt]
\multicolumn{16}{l}{\textbf{\textit{Principle 2 (RQ2, Communication State): SharedMemory boosts reasoning}} \textit{(claim direction $+$)}} \\
 & SM present & 13 & 10 & 0 & $+0.072$ & .002 & 10 & 6 & 0 & $+0.135$ & .011 & 9/10 & $<$.001 & 9/10 & .011 \\
 & SM 1st & 13 & 9 & 0 & $+0.072$ & .002 & 8 & 4 & 0 & $+0.143$ & .030 & 7/8 & $<$.001 & 7/8 & .035 \\
\addlinespace[2pt]
\multicolumn{16}{l}{\textbf{\textit{Principle 3 (RQ3, Sub-Topological Prompts): Richer prompt structure helps}} \textit{(claim direction $+$)}} \\
 & mean chars & 13 & 11 & 0 & $+0.073$ & .002 & 13 & 6 & 0 & $+0.118$ & .002 & 13/13 & $<$.001 & 13/13 & $<$.001 \\
 & max chars & 13 & 10 & 0 & $+0.074$ & .002 & 13 & 7 & 0 & $+0.112$ & .002 & 13/13 & $<$.001 & 13/13 & $<$.001 \\
 & \#placeh. & 13 & 9 & 0 & $+0.102$ & .002 & 13 & 7 & 0 & $+0.059$ & .003 & 11/13 & $<$.001 & 11/13 & .011 \\
\bottomrule
\end{tabular}
\caption{\textbf{Statistical Summary of Mined Design Principles.} Per variant: gate-admitted puzzles ($n$), cells with strictly positive ($k_+$) / negative ($k_-$) 95\% cluster-Wald CIs, median effect size ($\Delta$), and Wilcoxon $p$-values ($p_{\mathrm{W}}$) on train and held-out test splits, with out-of-sample sign agreement ($p_{\mathrm{B}}^{\cap}$) and conditional test verification ($p_{\mathrm{B}}^{\mathrm{tr}}$).}
\label{tab:laws-headline}
\end{table*}

We first profile the corpus (coverage, sampling sufficiency, and gains over single-agent baselines; \S~\ref{subsec:apexdb}) and then present the design principles mined and verified at three levels of granularity (\S~\ref{subsec:analysis}).

\subsection{Workflow Corpus and Coverage}
\label{subsec:apexdb}

Our corpus comprises $28{,}568$ executable multi-agent workflows across $26$ \enigmata{} puzzles ($1{,}701$ completed \aflow{} search traces out of $1{,}740$ grid cells; Tables~\ref{tab:dataset_structure} and \ref{tab:enigmata_score_distribution}, \S~\ref{apdx:add_res}), evaluated on 20-instance bundles at $\$0.0064$ per puzzle instance ($\sim\$4{,}454$ total compute; Table~\ref{tab:cost_per_instance}, \S~\ref{apdx:cost}). 
Workflows exercise all six operators individually and in chains: \textsc{Sequential} ($47.2\%$), \textsc{Star} ($44.5\%$), \textsc{Debate} ($22.3\%$), \textsc{SharedMemory} ($20.8\%$), \textsc{DebateWithModerator} ($11.8\%$), and \textsc{StarWithCoordinator} ($8.8\%$; Tables~\ref{tab:operator_percentages}--\ref{tab:most_common_operator_combinations}).

\begin{figure}[htbp]
    \centering
    \includegraphics[width=0.9\linewidth]{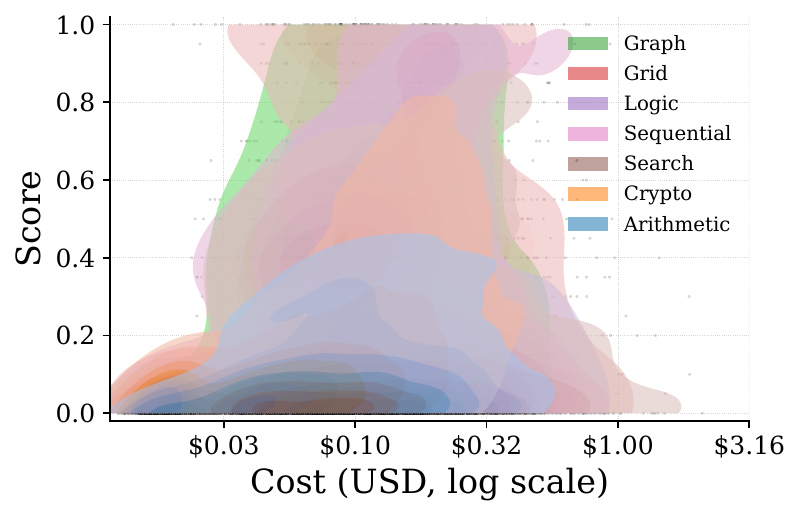}
    \caption{\textbf{Design-Space Coverage.} Joint distribution of workflow evaluation accuracy versus inference cost (USD, log scale) across task categories, spanning a broad spectrum of accuracy-cost trade-offs.}
    \label{fig:coverage}
\end{figure}

\begin{figure}[htbp]
    \centering
    \includegraphics[width=0.9\linewidth]{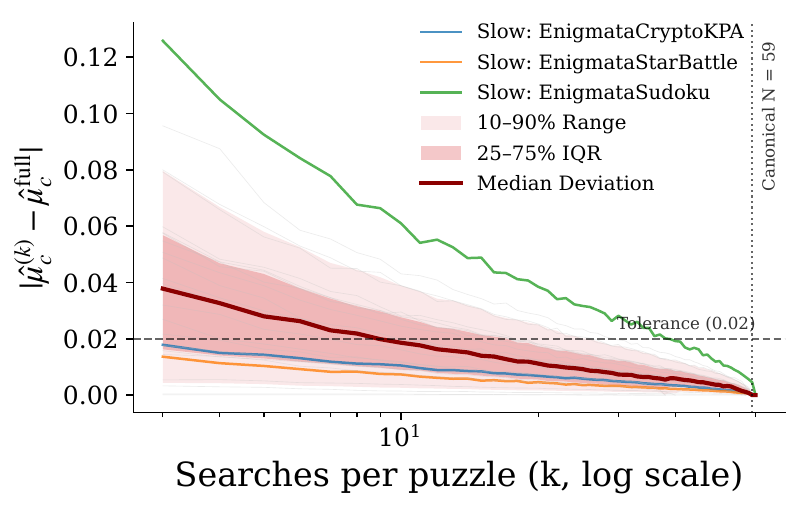}
    \caption{\textbf{Sampling Sufficiency.} Absolute deviation of the running cell mean from its full-sweep value across 26 puzzles as a function of search count $k$. Rapid convergence below $0.02$ confirms estimator stability.}
    \label{fig:sufficiency}
\end{figure}

\begin{figure}[htbp]
    \centering
    \includegraphics[width=\linewidth]{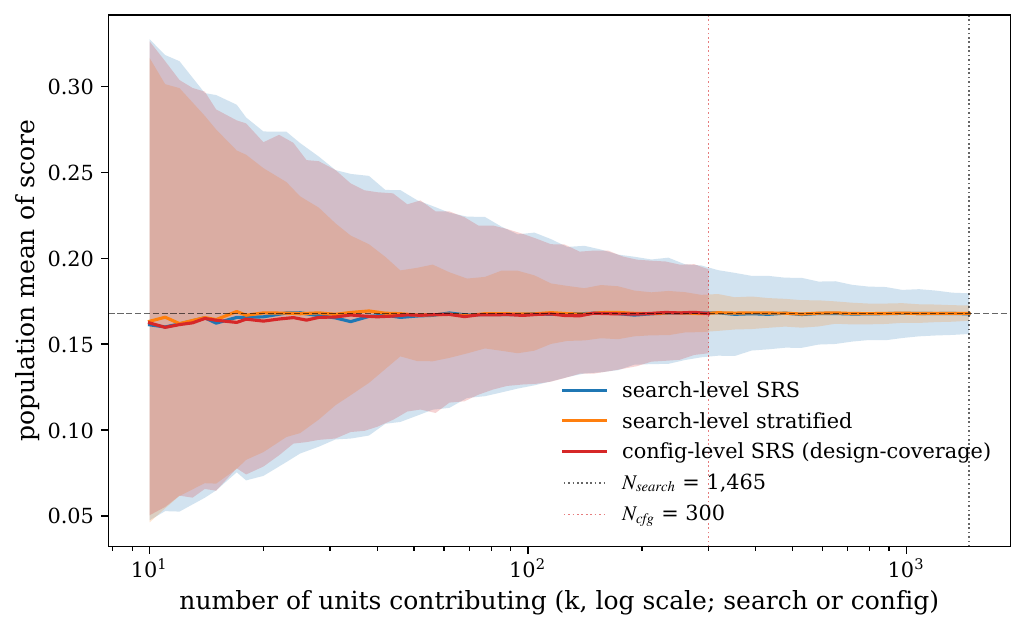}
    \caption{Convergence of the population-mean score estimator. For each $k$ (log scale),
  we draw $B{=}2{,}000$ bootstrap replicates of $k$ resampling units with                                                            
  replacement and plot the bootstrap median (line) and central $95\%$ band of                                                        
  $\hat{\mu}(k)=\mathrm{mean}_{\mathrm{rows}}(\mathrm{score})$. Three resampling                                                     
  units: SRS over searches (blue), SRS stratified by (dataset, difficulty)                                                           
  (orange), SRS over configurations (red). Dashed horizontal line:
  full-population mean. Dotted verticals: $N_{\mathrm{search}}$ (black),                                                             
  $N_{\mathrm{cfg}}$ (red). Cleaned subset (4 sliding-tile puzzles dropped):
  $1{,}465$ searches, $300$ configurations, $26$ puzzles,                                                                            
  $\hat{\mu}_{\mathrm{full}}{=}0.168$.}
    \label{fig:population_mean}
\end{figure}

\paragraph{Single-Agent Baselines, Coverage, and Sufficiency}
Against one-shot evaluations of the single-agent backbones (\texttt{GPT-5.4}, \texttt{GPT-5.4-mini}, \texttt{GPT-5.4-nano}) on the identical puzzle bundles (Table~\ref{tab:baselines}, \S~\ref{app:baseline}), multi-agent workflows using only \texttt{mini}/\texttt{nano} workers substantially outperform even the stronger \texttt{GPT-5.4} baseline on tasks requiring iterative constraint propagation, lifting accuracy from $0.100 \to 0.891$ (mean) and $1.000$ (max) on NaturalLanguageNavigation, $0.033 \to 0.359$ on MagicSquare, and $0.467 \to 0.849$ on StackPermutation. 
Full MCTS traces capture workflows spanning three orders of magnitude in cost ($\$0.01$ to $>\$3.00$ per bundle) and the full $[0, 1]$ accuracy range (Fig.~\ref{fig:coverage}), while the running cell mean converges within a $0.02$ tolerance well before the full budget ($N_{\text{cfg}} = 59$ searches per puzzle; Fig.~\ref{fig:sufficiency}, Fig.~\ref{fig:population_mean}).

\subsection{Mining and Verifying Design Principles}
\label{subsec:analysis}

\paragraph{Candidate Pool and Predictability Gating}
We executed $10$ main \autoresearch{} runs (\texttt{graph.py} only vs.\ \texttt{graph.py} $+$ \texttt{prompt.py}; Fig.~\ref{fig:autoresearch_trace}), totaling $300$ automated feature-engineering trials (\S~\ref{app:add_laws}). 
While \texttt{graph.py}-only feature sets plateaued at validation $R^2 = 0.521$ ($8$ features), jointly parsing \texttt{prompt.py} lifted validation $R^2$ to $0.638$ ($27$ features), yielding $70$ candidate binarized contrasts. 
Within each puzzle, workflows are split into high and low arms as defined in \S~\ref{subsec:formulate} ($\ge 20$ workflows per arm) to estimate $\Delta$ with $95\%$ cluster-robust Wald CIs clustered by \aflow{} search ID (\S~\ref{app:cluster-robust}). 
To avoid diluting contrasts with floor/ceiling tasks, a pre-test \emph{predictability gate} admits a puzzle iff a LightGBM surrogate trained on $D_{\text{train}}$ achieves validation $R^2 > 0.05$ and Spearman $\rho > 0.30$ on $D_{\text{val}}$ (\S~\ref{app:gate-calibration}). 
Calibrated against a permutation null ($R^2_{p95} \le -0.06$, $\rho_{p95} \le 0.23$; Fig.~\ref{fig:gate_forest}), this gate admits $13$ predictable puzzles prior to accessing $D_{\text{test}}$ (Fig.~\ref{fig:effect}, Table~\ref{tab:laws-headline}; Table~\ref{tab:validation-ladder}):

\paragraph{Principle 1 (\textit{RQ1, Entry-Point Topology}): \textsc{Star}-first fan-out degrades reasoning}
When a workflow begins with an isolated \textsc{Star} operator (parallel workers without lateral communication followed by one-shot aggregation), accuracy drops relative to alternative entry topologies on $7/13$ train puzzles ($\text{median }\Delta = -0.036$, $p_{\mathrm{W}} = .008$), with $0/13$ strictly positive cells (Fig.~\ref{fig:effect} left; Table~\ref{tab:law-2}). 
This negative effect persists on the held-out test split ($\text{median }\Delta = -0.045$, $p_{\mathrm{W}} = .038$, $9/10$ conditional sign replication), corroborating \citet{scalingagent}: uncoordinated parallel fan-out amplifies reasoning errors that an aggregator must then reconcile without peer verification.

\paragraph{Principle 2 (\textit{RQ2, Communication State}): \textsc{SharedMemory} outperforms conversational dialogue}
Whereas LLM frameworks default to conversational message passing (\textsc{Debate}), workflows using blackboard-style \textsc{SharedMemory} \citep{hayes1985blackboard}, in which agents sequentially read and append to a persistent written document, outperform workflows without it on $10/13$ train puzzles ($\text{median }\Delta = +0.072$, $p_{\mathrm{W}} = .002$) and $6/10$ test puzzles ($\text{median }\Delta = +0.135$, $p_{\mathrm{W}} = .011$, $9/10$ sign agreement; Fig.~\ref{fig:effect} middle, Table~\ref{tab:law-1}), achieving the highest corpus-wide mean accuracy of all six operators ($0.23$ vs.\ $0.18$ for \textsc{Debate} and $0.16$ for \textsc{Star}; Table~\ref{tab:operator_percentages}), showing that persistent written state mitigates ``lossy communication'' in multi-round dialogue \citep{scalingagent}.

\paragraph{Principle 3 (\textit{RQ3, Sub-Topological Prompt Structure}): prompt specificity and placeholder routing predict accuracy}
Below the categorical topology grain, workflows with above-median prompt template character length (\texttt{mean chars} in \texttt{prompt.py}) achieve strictly positive accuracy gains on $11/13$ train ($\text{median }\Delta = +0.073$, $p_{\mathrm{W}} = .002$) and $6/13$ test puzzles ($\text{median }\Delta = +0.118$, $p_{\mathrm{W}} = .002$, $13/13$ sign agreement; Fig.~\ref{fig:effect} right, Table~\ref{tab:law-3}). 
Similarly, the count of \texttt{\{name\}} placeholders (\texttt{\#placeh.}) replicates this positive effect on $9/13$ train ($\text{median }\Delta = +0.102$) and $7/13$ test puzzles ($\text{median }\Delta = +0.059$, $11/13$ sign agreement). 
We do not claim character length is causal in isolation; rather, sub-topological prompt specification and context routing carry effect sizes ($\Delta \approx +0.06\text{ to }+0.12$) on par with macro-topology selection, which is why \texttt{prompt.py} features lift validation $R^2$ from $0.521$ to $0.638$ (Fig.~\ref{fig:autoresearch_trace}).

\paragraph{Out-of-Sample Verification and Test Precision}
Across all seven variants in Table~\ref{tab:laws-headline}, two-sided Wilcoxon tests ($p_{\mathrm{W}} < 0.05$), binomial sign-agreement tests ($p_{\mathrm{B}}^{\cap} \le .001$), and conditional test-verification tests ($p_{\mathrm{B}}^{\mathrm{tr}} \le .035$) confirm directional reproducibility on $D_{\text{test}}$. 
Median effect sizes hold on test, but fewer puzzles clear strict 95\% cluster-Wald CIs ($k_+ / k_-$) than on train (e.g., $4/12$ vs.\ $7/13$ for \textsc{Star}-first; $6/10$ vs.\ $10/13$ for \textsc{SharedMemory}) because the $15.6\%$ test split has fewer search clusters ($G \in [4, 14]$ vs.\ $[19, 47]$; \S~\ref{app:cluster-robust}): the findings are directionally robust, with finite-sample uncertainty on task-specific magnitudes.

\section{Discussion and Future Directions}
\label{sec:discussion_limitations}

\paragraph{Full-Trace Mining vs.\ Point Optimization}
Automated workflow search and empirical design-space science are complementary: the non-argmax rounds that search discards provide exactly the negative and intermediate comparisons needed to explain why designs fail and which edits help. 
Full traces also expose accuracy-cost trade-offs: \textsc{SharedMemory} costs more than \textsc{Debate} or \textsc{Star} (\$0.25 vs.\ \$0.13; Table~\ref{tab:operator_percentages}) as the shared document grows, yet its persistent written state avoids the information loss of multi-round dialogue and yields the highest mean accuracy in the corpus.

\paragraph{Beyond Topology-Only Benchmarks}
Macro-topology alone (\texttt{graph.py}) explains only part of workflow variance ($R^2 = 0.521$); jointly mining sub-topological prompt structure (\texttt{prompt.py}) lifts out-of-sample predictive power to $R^2 = 0.638$ (\S~\ref{subsec:analysis}). 
Topology benchmarks should therefore report and control for prompt template richness and placeholder routing rather than treating prompts as fixed constants: an ostensibly superior communication graph can be bottlenecked by underspecified role instructions or missing context placeholders.

\paragraph{Operator Evolution and Tool-Rich Domains}
Feeding mined regularities back into the operator registry could guide the synthesis of new operator primitives \citep{novikov2025alphaevolve,genesys} or prune suboptimal entry-point patterns (such as uncoordinated \textsc{Star}-first fan-out) from the optimizer's search prior, and extending the pipeline to tool-rich domains such as coding \citep{jimenez2023swe}, games \citep{minedojo}, and social simulations \citep{sociodojo} will test how topological and prompt principles interact with external tool use and environment feedback.

\section{Conclusion}

We presented an end-to-end pipeline that moves multi-agent LLM workflow design from manual ``folk wisdom'' toward systematic statistical mining of design principles. Full-trace \aflow{} search over 26 deterministically verified \enigmata{} tasks yielded $28{,}568$ executable workflows, and a sandboxed, search-level-split \autoresearch{} loop over 300 trials surfaced out-of-sample verified principles at three granularities: \textsc{Star}-first fan-out degrades accuracy, blackboard-style \textsc{SharedMemory} outperforms conversational debate, and prompt template specificity and placeholder routing matter as much as macroscopic topology. 
We release the workflow corpus, evaluation harness, and sandboxed analysis pipeline for reproducible empirical research on multi-agent systems.

\section*{Limitations}

To isolate multi-agent coordination and prompt structure from external confounds, our experiments focus on tool-free, deterministically verified reasoning tasks from \enigmata{} and evaluate workflows within a single model family (\texttt{GPT-5.4} as the MCTS optimizer and \{\texttt{GPT-5.4-mini}, \texttt{GPT-5.4-nano}\} as worker and aggregator backbones; \S~\ref{apdx:exp}). 
While this controlled setup eliminates grading noise and API variance (Appendix~\ref{app:theory}), extending our pipeline to open-weight model families, alternative workflow search algorithms, and tool-interactive domains (such as repository-level coding or web navigation) is a natural direction for future work. 
Finally, because our corpus harvests trajectories generated along MCTS optimization runs, the mined principles reflect structural regularities within the space of executable workflows explored by the optimizer rather than unconstrained random code generation.

\bibliographystyle{plainnat}
\bibliography{references}

\clearpage
\appendices
\section{Methodological Formalization and Statistical Estimands}
\label{app:theory}

As introduced in \S~\ref{sec:framework}, our pipeline mines empirical design principles from observational workflow trajectories $\mathcal{D} = \{(W_i, T_i, \bm_i)\}_{i=1}^{|\mathcal{D}|}$. 
Here we formalize the statistical estimands recovered by our approach and clarify how each stage (\aflow{} traversal, \enigmata{} verification, and search-level sandboxed \autoresearch{}) addresses key threats to validity when learning from non-randomized workflow corpora; Table~\ref{tab:bias-component-map} maps each challenge to the pipeline component that addresses it.

\begin{table*}[t]
\centering
\footnotesize
\setlength{\tabcolsep}{6pt}
\renewcommand{\arraystretch}{1.05}
\begin{tabular}{@{}>{\raggedright\arraybackslash}p{2.3cm} >{\raggedright\arraybackslash}p{4.8cm} >{\raggedright\arraybackslash}p{2.6cm} >{\raggedright\arraybackslash}p{5.0cm}@{}}
\toprule
\textbf{Challenge} & \textbf{Manifestation in MAS Evaluation} & \textbf{Pipeline Stage} & \textbf{Methodological Design} \\
\midrule
Design-space coverage \& implementation variance
& Manual studies test only a few fixed topologies; argmax search discards intermediate designs; naive random code generation produces broken prompts or syntax errors.
& \aflow{} Search \newline (\emph{Sampling};~\S~\ref{subsec:traverser})
& Retains \emph{all} $N=20$ MCTS rounds across $K=5$ runs over 6 composable operators, sampling both failed and high-performing workflows under a consistent LLM optimizer policy (\S~\ref{app:theory}). \\
\addlinespace[2pt]
Domain \& evaluation confounds
& Downstream benchmarks conflate coordination topology with external tool APIs, domain knowledge, or noisy LLM-as-a-judge grading.
& \enigmata{} Verification \newline (\emph{Sampling};~\S~\ref{subsec:evaluation})
& Tool-free evaluation on 26 parametric abstract-reasoning tasks with exact, deterministic rule-based verifiers and fixed 20-instance bundles per difficulty tier (\S~\ref{app:enigmata}). \\
\addlinespace[2pt]
High-dimensional feature overfitting \& leakage
& Workflow code (\texttt{graph.py}) and prompts (\texttt{prompt.py}) admit infinitely many features; manual selection is narrow, while unconstrained search overfits or leaks test data.
& \autoresearch{} Mining \newline (\emph{Analysis};~\S~\ref{subsec:autoresearch})
& Sandboxed AST/prompt feature mining with an $L_0$-penalized validation $R^2$ objective on strict search-level splits; pre-specified directional verification on held-out test searches (\S~\ref{apdx:methodology}). \\
\bottomrule
\end{tabular}
\caption{\textbf{Core methodological challenges in empirical MAS design-space mining and corresponding pipeline components.} Each stage of the pipeline addresses a specific operational threat to extracting reproducible design principles. }
\label{tab:bias-component-map}
\end{table*}

\subsection{Implementation Variance and Policy-Conditioned Sampling}

When mapping a raw workflow $W$ (consisting of \texttt{graph.py} and \texttt{prompt.py}) into an interpretable feature vector $\ba = \phi(W)$, microscopic implementation details $\theta_s \in \Theta_s$ (such as exact phrasing choices or variable names not captured by $\phi$) remain unmodeled. 
Marginalizing over the distribution of implementation details under a sampling policy $\mathcal{L}$, the expected performance of workflows with feature profile $\ba$ on task $\bt$ is:
\begin{equation}
    \resizebox{0.88\columnwidth}{!}{$\displaystyle \mu_{\mathcal{L}}(\ba, \bt) = \int_{\Theta_s} \mathbb{E}[\bm \mid \ba, \bt, \theta_s] \, P_{\mathcal{L}}(\theta_s \mid \ba, \bt) \, d\theta_s$}
    \label{eq:backdoor}
\end{equation}

\paragraph{Why MCTS Trajectory Harvesting Beats Naive Random Code Sampling}
If $\mathcal{L}$ were a uniform random code-and-string generator over $\Theta_s$, the vast majority of sampled workflows would contain syntax errors or incoherent prompts, causing extreme variance in $\theta_s$ to swamp structural differences in $\ba$. 
Conversely, retaining only the final argmax $W^*$ from an optimizer collapses support onto a single point per search. 
By harvesting all $N = 20$ MCTS rounds generated under a fixed LLM optimizer (\texttt{GPT-5.4} at temperature $0$) with a mixed uniform/softmax parent-selection policy ($\lambda = 0.2$; Eq.~\ref{eq:mixture}), \aflow{} restricts $\Theta_s$ to syntactically valid, competently prompted implementations while preserving broad variation across operators and prompt templates. 
Importantly, as noted in our Limitations, $\mu_{\mathcal{L}}(\ba, \bt)$ represents a \emph{policy-conditioned observational estimand} relative to the foundation model's capability frontier rather than an unconditioned interventional effect.

\subsection{Evaluation Variance and Fixed-Bundle Verification}

Let $G(W, T)$ denote the observed evaluation score of workflow $W$ on task bundle $T = \{x_1, \dots, x_M\}$. 
In open-ended benchmarks that rely on external tool APIs, web state, or LLM-as-a-judge grading, the observed score decomposes as:
\begin{equation}
    \widehat{\bm}(W, T) = \mu_{\text{reasoning}}(W, T) + \xi_{\text{domain/tool}} + \epsilon_{\text{judge}}
\end{equation}
where $\xi_{\text{domain/tool}}$ reflects external API or domain-knowledge bottlenecks and $\epsilon_{\text{judge}}$ represents grading noise.

\paragraph{Deterministic Verification on Reused Bundles (\enigmata{})}
Evaluating tool-free workflows on \enigmata{} \citep{enigmata} eliminates subjective judge variance ($\epsilon_{\text{judge}} = 0$) because every puzzle instance is scored by an exact, deterministic rule-based verifier. 
Furthermore, because the same fixed bundle of $M = 20$ deterministically generated instances is reused across all $K = 5$ searches and $N = 20$ MCTS rounds within each (puzzle, difficulty) cell (\S~\ref{app:enigmata}), instance-sampling variance between workflows compared within the same cell cancels out in paired per-puzzle contrasts $\Delta = \mathbb{E}[\text{score} \mid \text{high arm}] - \mathbb{E}[\text{score} \mid \text{low arm}]$.

\subsection{Overfitting, Trajectory Leakage, and Out-of-Sample Verification}

Let $S(\phi) \in \{0, 1\}$ denote the indicator that a candidate feature extractor $\phi \in \mathbf{H}$ is selected during feature mining. 
Two forms of overfitting threaten automated feature discovery over workflow code: 
(1)~\emph{within-search trajectory leakage}, where correlated MCTS rounds from the same search straddle training and evaluation splits; and 
(2)~\emph{test-set selection leakage}, where feature selection $S(\phi)$ conditions on held-out test outcomes $\bm_{\text{test}}$.

\paragraph{Search-Level Splitting and Pre-Specified Test Verification}
Our analysis phase prevents both failure modes structurally. 
First, partitioning $\mathcal{D}$ into $D_{\text{train}}$, $D_{\text{val}}$, and $D_{\text{test}}$ at the \emph{\aflow{} search level} guarantees that all $N = 20$ rounds of any search trajectory reside in a single split. 
Second, the \autoresearch{} loop evaluates the $L_0$-penalized objective $\mathcal{F}(\phi, \psi, \kappa)$ (Eq.~\ref{eq:srm}) exclusively on $D_{\text{val}}$ inside an isolated sandbox with no access to $D_{\text{test}}$ or row/round identifiers. 
Because both the candidate features $\phi$ and the hypothesized contrast directions are fixed prior to accessing $D_{\text{test}}$, the selection mechanism satisfies conditional independence $S(\phi) \perp\!\!\!\perp \bm_{\text{test}} \mid \ba$, ensuring that test-split contrasts provide a valid, uncontaminated out-of-sample replication check.

\section{Experiment Details and Hyperparameters}
\label{apdx:exp}
This section consolidates the configuration choices that govern the sampling and analysis phases. Hyperparameters are grouped by component, model selection, workflow constraints, the \aflow{} search, and the predictability gate (Table~\ref{tab:hyperparameters}). Table~\ref{tab:model_snapshots} pins the OpenAI model snapshots used throughout.

\begin{table}[htbp]
    \centering
    \resizebox{\columnwidth}{!}{%
    \begin{tabular}{cc}
        \toprule[1.5pt]
         \textbf{Hyperparameters} & \textbf{Values} \\
         \midrule
         \multicolumn{2}{c}{Models} \\
         \midrule
         Optimizer &  GPT-5.4 \\
         Opt. Temperature (\temp) & 0 \\
         Worker &  [GPT-5.4-mini, GPT-5.4-nano] \\
         Worker temperature (\temp) &  0 \\
         Aggregator & [GPT-5.4-mini, GPT-5.4-nano] \\
         Aggregator temperature (\temp) &  0 \\
         \midrule
         \multicolumn{2}{c}{Workflow Constraints} \\
         \midrule
         Maximal number of calls per workflows &  20 \\
        \# Operators &  6 \\
         \midrule
         \multicolumn{2}{c}{\aflow{}} \\
         \midrule
         Number of evaluation step per round &  1 \\
         Number of search steps &  20 \\
         Number of searches per experimental condition &  5 \\
         Number of difficulty bundles & 2,3 \\
         Number of puzzles per bundle & 20 \\ 
         Exploration/Exploitation weight ($\lambda$) & 0.2 \\ 
         Top-$k$ score-based selection & 4 \\ 
         Score influence factor $\lambda$ & 0.4 \\ 
         \midrule
         \multicolumn{2}{c}{Predictability gating} \\
         \midrule
         $R^2$ threshold  & $0.05$ \\
         $\rho$ threshold & $0.30$ \\
         Min entry per cell & $20$ \\
         \midrule
         Model & LightGBM \\
         Learning rate & $0.05$ \\
         Number of leaves & $31$ \\
         Min child samples & $10$ \\
         \bottomrule[1.5pt]
    \end{tabular}%
    }
    \vspace{5pt}
    \caption{Hyper-parameters}
    \label{tab:hyperparameters}
\end{table}

\paragraph{Model selection}
The Optimizer role, which proposes structural workflow modifications during \aflow{} search, is fixed to GPT-5.4 in order to maintain a single, stationary generative policy across the corpus (cf.\ \S~\ref{app:theory}). The Worker and Aggregator roles are drawn from $\{\text{GPT-5.4-mini}, \text{GPT-5.4-nano}\}$, providing a controlled $2 \times 2$ contrast between worker and aggregator capability while keeping the per-search call budget tractable.

\paragraph{Workflow constraints}
Each generated workflow is capped at $20$ atomic LLM calls, which bounds the per-evaluation cost while still admitting compositions of multiple operators (\S\ref{apdx:operator_set}). Because each workflow is evaluated on the full $20$-instance bundle, this caps the per-search call budget at $20 \times 20 = 400$ atomic calls per round and at most $8{,}000$ calls for an entire \aflow{} search trace. The operator set comprises the six primitives listed in Table~\ref{tab:operators}.
 
\paragraph{\aflow{} search}
Each (puzzle, bundle, model) cell is searched $K = 5$ times for $N = 20$ MCTS rounds. The selection mixture (Eq.~\ref{eq:mixture}) uses $\lambda = 0.2$ to balance uniform exploration against score-weighted exploitation, with score temperature $\alpha = 0.4$ and a top-$k$ pool of size $k = 4$.
 
\paragraph{Predictability gate}
The LightGBM surrogate (200 trees, learning rate $0.05$, $31$ leaves, \texttt{min\_child\_samples}~$=10$, fixed seed) used for the predictability gate admits a puzzle if $R^2 > 0.05$ \emph{and} Spearman $\rho > 0.30$ on its validation rows; the calibration of these thresholds against a permutation null is detailed in \S\ref{app:gate-calibration}. The minimum cell occupancy of $20$ workflows in either arm is enforced for every per-puzzle binarized contrast.

\begin{table}[htbp]
    \centering
    \begin{tabular}{cc}
        \toprule
         Model name & Snapshot \\
         \midrule
         GPT-5.4 & gpt-5.4-2026-03-05 \\
         GPT-5.4-mini & gpt-5.4-mini-2026-03-17 \\
         GPT-5.4-nano & gpt-5.4-nano-2026-03-17 \\
         \bottomrule
    \end{tabular}
    \vspace{5pt}
    \caption{Model snapshots for the OpenAI models.}
    \label{tab:model_snapshots}
\end{table}

\subsection{Enigmata puzzles}

We selected a subset of the \enigmata{} puzzles, namely we focused on 30 puzzles for which the code-base provided both generator and verifiers, these puzzle collection covers 2 Arithmetic, 2 Crypto, 4 Graph, 8 Grid, 1 Logic, 6 Search and 7 Sequential puzzles (Table~\ref{tab:enigmata_score_distribution}).
\paragraph{Puzzle Search space}
Our sampling grid varies seven factors: (1) puzzle type, (2) difficulty bundle, (3) optimizer model, (4) worker model, (5) aggregator model, (6) \aflow{} search run, and (7) MCTS iteration. Detailed information about these factors are available in Table~\ref{tab:hyperparameters}.

\begin{enumerate}
    \item \textbf{Puzzle type:} $\symPuz = \valPuz$ \enigmata{} puzzles, e.g., Sudoku, TicTacToe, ZebraLogic.
    \item \textbf{Difficulty bundles:}  \emph{easy}/\emph{medium}/\emph{hard}, where 27 puzzles expose all three ($\symBun = 3$) and 3 puzzles, Hitori, LightUp, TicTacToe, only \emph{medium}/\emph{hard} ($\symBun = 2$). A bundle is a collection of puzzle instance associated with these difficulties.
    \item \textbf{Optimizer model:} $\symOpt = \valOpt$ the LLM backbone that proposes workflow updates in \aflow{}, e.g., GPT-5.4.
    \item \textbf{Worker model:}~$\symWrk = \valWrk$, LLM backbones running individual task work, the Star-operator leaves.
    \item \textbf{Aggregator model:}~$\symAgg = \valAgg$, LLM backbones handling operator-level synthesis.
    \item \textbf{\aflow{} search run:}~$\symSeed = \valSeed$ \aflow{} searches per (puzzle, bundle, model) cell.
    \item \textbf{\aflow{} search execution:} $\symMcts = \valMcts$ MCTS rounds per \aflow{} search.
\end{enumerate}

For each (puzzle, bundle) pair we generate $\valInst$ puzzle instances deterministically from their associated \enigmata{} difficulty level. 

A naive grid with uniform $\symBun = 3$ would yield $\valPuz \cdot 3 \cdot \valOpt \cdot \valWrk \cdot \valAgg \cdot \valSeed \cdot \valMcts = 36{,}000$ workflows from $1{,}800$ searches. Nonetheless, given the three puzzles missing \emph{easy} there are $27 \cdot 3 + 3 \cdot 2 = 87$ (puzzle, bundle) cells and $87 \cdot \valOpt \cdot \valWrk \cdot \valAgg \cdot \valSeed = 1{,}740$ searches, leading to at most $34{,}800$ workflows. Out of the possible $1{,}740$ searches, $1{,}701$ completed successfully (37 mid-search aborts and 2 dispatcher misses, uniformly spread across the grid), producing $1{,}701 \cdot \valMcts = 34{,}020$ ($97.7\%$) candidate workflows.  

\paragraph{Executable workflow filtering} From these $34{,}020$ workflows we further filter $1{,}031$ (3\%) for which the graphs are either missing or could not be evaluated, for instance because a custom prompt generated by the \aflow{} Optimizer model was undefined or mis-referenced.

\paragraph{Call-budget constraint}
Each workflow is authorized a maximal number of atomic LLM calls of $20$, given the type of puzzles, this still allows workflows that contain multiple operators with few calls or operators that leverage single operators with many calls. Given that each workflow is evaluated on all the puzzles of a bundle, this leads to $20\times20$ calls per \aflow{} search round, and $8{,}000$ calls at most for a single \aflow{} search.

\paragraph{Consistent Sequential Puzzle failures}
After further analyses, we removed 4 of the Sequential puzzles: EightPuzzle, NinePuzzle, FifteenPuzzle, SixteenPuzzle.
Their minimal and maximal scores remained  at 0 across all searches, from further inspection the LLM answers were either wrong attempts or simply stated that there was no solution to the puzzles.
Following this removal step we obtain a reduced set of $28{,}568$ workflows.

\subsubsection{Feature-Mining Search Evolution}
\label{subsubsec:autoresearch_evolution}
We conduct multiple \autoresearch{} runs with varying prompts, aiming at optimizing the validation set $R^2$ while keeping the number of features small.
We considered both validation $R^2$ over the entire population, and average $R^2$ reported across per-puzzle split.
These searches keep track of the evolution of the validation $R^2$, features, and size of the feature set, allowing us to restore a search step and leverage discovered feature and combination of features, Fig.~\ref{fig:autoresearch_trace} shows an example of such traces.

\subsection{Analysis Phase}

Each candidate design principle is evaluated per puzzle. For a given feature, workflows are partitioned within each puzzle into a \emph{high arm} and a \emph{low arm}: for binary features (e.g., presence of a SharedMemory operator) the high arm consists of workflows where the feature is present and the low arm those where it is absent; for count or continuous features (e.g., mean prompt-template character length) the partition is the per-puzzle median split, with the high arm strictly above the median and the low arm at or below. The per-puzzle effect size is the difference in mean score between the two arms, $\Delta = \mathbb{E}[\text{score} \mid \text{high arm}] - \mathbb{E}[\text{score} \mid \text{low arm}]$, and per-puzzle verdicts are aggregated to a principle-level verdict.

\paragraph{Predictability gate} For each puzzle, we fit a LightGBM regressor on the training-split workflows (200 trees, learning rate 0.05, 31 leaves, \texttt{min\_child\_samples} = 10, fixed seed) predicting score from design features, and evaluate it on the puzzle's held-out validation workflows.
A puzzle is admitted iff validation $R^2 > 0.05$ \emph{and} Spearman $\rho > 0.30$.
Both are required: a constant predictor attains $R^2 = 0$ with undefined $\rho$, while a rank-preserving but magnitude-compressing predictor attains $\rho > 0$ with $R^2 \leq 0$, so the conjunction enforces both calibration and ranking (see \S~\ref{app:gate-calibration}).

\paragraph{Feature-set for the predictability gate}
The predictability gate is parameterized by the feature set supplied to the regression surrogate, since the admission of a puzzle is meaningful only if the surrogate has access to features of the type the candidate principle concerns.
A puzzle whose validation $R^2$ exceeds the gate threshold under a feature set that includes prompt-length statistics provides no evidence that its score variation is explained by structural design choices; conversely, restricting the surrogate to structural features alone risks excluding puzzles whose design signal is real but weaker than a prompt signal.

We build the feature set using both handcrafted features that leverage the actual execution code of the workflows, associated external prompt files, and features that were extracted during \autoresearch{} explorations.

\subsubsection{Per-Puzzle Breakdowns of Mined Principles}

Tables~\ref{tab:law-2}--\ref{tab:law-3} break down the headline results of Table~\ref{tab:laws-headline} by puzzle, one table per principle. Each row gives the contrast $\Delta$ of a gate-admitted puzzle on the train (tr) and held-out test (te) splits; a dash marks a cell that is not testable on that split (fewer than $20$ workflows in an arm). The bottom rows report how many testable cells have a 95\% cluster-Wald CI strictly above ($k_+/n$) or below ($k_-/n$) zero, and the median $\Delta$.

\begin{table}[t]
\centering
\small
\setlength{\tabcolsep}{4pt}
\renewcommand{\arraystretch}{0.95}
\caption{Principle 1 (RQ1): \textsc{Star} as the first operator lowers score. Per-puzzle effect size $\Delta$ score (high arm $-$ low arm) on the train and held-out test splits. Feature definitions: \textbf{Star 1st} = binary indicator: 1 iff the first operator instantiated in \texttt{graph.py} (parsed in source order) is \texttt{Star}; high arm = 1, low arm = 0. $k_{+}/n$ and $k_{-}/n$ count cells whose 95\% cluster-Wald CI is strictly above / below zero (out of testable cells). Headline: Star 1st $-7/13$ on train (median $\Delta=-0.036$); Star 1st $-4/12$ on test (median $\Delta=-0.045$).}
\vspace{5pt}
\label{tab:law-2}
\begin{tabular}{l r r}
\toprule
 & \multicolumn{2}{c}{Star 1st} \\
\cmidrule(lr){2-3}
Puzzle & tr & te \\
\midrule
Campsite & $-0.095$ & $-0.111$ \\
CarPainting & $+0.004$ & $-0.021$ \\
CryptoKKA & $-0.052$ & $-0.076$ \\
CryptoKPA & $-0.012$ & $-0.003$ \\
Game24 & $-0.036$ & $-0.020$ \\
HamiltonianPath & $-0.037$ & $-0.032$ \\
Kakurasu & $-0.088$ & $-0.105$ \\
MagicSquare & $-0.070$ & $-0.012$ \\
StackPermutation & $-0.104$ & $-0.069$ \\
Sudoku & $+0.013$ & $-0.058$ \\
Sudoku2 & $-0.000$ & $-0.064$ \\
SumSkyscraper & $-0.026$ & -- \\
ZebraLogic & $-0.005$ & $+0.147$ \\
\midrule
$k_{+}/n$ & 0/13 & 0/12 \\
$k_{-}/n$ & 7/13 & 4/12 \\
median $\Delta$ & $-0.036$ & $-0.045$ \\
\bottomrule
\end{tabular}
\end{table}

\begin{table}[t]
\centering
\small
\setlength{\tabcolsep}{4pt}
\renewcommand{\arraystretch}{0.95}
\caption{Principle 2 (RQ2): \textsc{SharedMemory} operators raise score. Per-puzzle effect size $\Delta$ score (high arm $-$ low arm) on the train and held-out test splits. Feature definitions: \textbf{SM present} = count of SharedMemory operator instantiations in \texttt{graph.py}, median-split within each puzzle; the per-puzzle median is 0 for every puzzle so the high arm = workflows with $\geq$1 SharedMemory operator and the low arm = workflows with none. $k_{+}/n$ and $k_{-}/n$ count cells whose 95\% cluster-Wald CI is strictly above / below zero (out of testable cells). Headline: SM present $+10/13$ on train (median $\Delta=+0.072$); SM present $+6/10$ on test (median $\Delta=+0.135$).}
\vspace{5pt}
\label{tab:law-1}
\begin{tabular}{l r r}
\toprule
 & \multicolumn{2}{c}{SM present} \\
\cmidrule(lr){2-3}
Puzzle & tr & te \\
\midrule
Campsite & $+0.168$ & $+0.178$ \\
CarPainting & $+0.003$ & $+0.076$ \\
CryptoKKA & $+0.178$ & $+0.139$ \\
CryptoKPA & $+0.036$ & $+0.009$ \\
Game24 & $+0.072$ & -- \\
HamiltonianPath & $+0.055$ & $+0.055$ \\
Kakurasu & $+0.282$ & $+0.131$ \\
MagicSquare & $+0.172$ & $+0.161$ \\
StackPermutation & $+0.273$ & $+0.294$ \\
Sudoku & $+0.010$ & -- \\
Sudoku2 & $+0.107$ & $+0.161$ \\
SumSkyscraper & $+0.068$ & -- \\
ZebraLogic & $+0.012$ & $-0.051$ \\
\midrule
$k_{+}/n$ & 10/13 & 6/10 \\
$k_{-}/n$ & 0/13 & 0/10 \\
median $\Delta$ & $+0.072$ & $+0.135$ \\
\bottomrule
\end{tabular}
\end{table}

\begin{table}[t]
\centering
\small
\setlength{\tabcolsep}{3.4pt}
\renewcommand{\arraystretch}{0.95}
\caption{Principle 3 (RQ3): Longer, structurally rich prompts raise score. Per-puzzle effect size $\Delta$ score (high arm $-$ low arm) on the train and held-out test splits. Feature definitions: \textbf{mean chars} = mean character length of the module-level string-assignment prompt templates in \texttt{prompt.py}, median-split within each puzzle (high arm = above the puzzle's median); \textbf{\#placeh.} = total occurrence count of \texttt{\{name\}}-style format placeholders (matched by \texttt{r"\textbackslash\{[a-zA-Z\_][a-zA-Z0-9\_]*\textbackslash\}"}, with repeats) across all module-level prompt-template strings in \texttt{prompt.py}, median-split within each puzzle. $k_{+}/n$ and $k_{-}/n$ count cells whose 95\% cluster-Wald CI is strictly above / below zero (out of testable cells). Headline: mean chars $+11/13$ on train (median $\Delta=+0.073$) and $+6/13$ on test (median $\Delta=+0.118$); \#placeh.\ $+9/13$ on train (median $\Delta=+0.102$) and $+7/13$ on test (median $\Delta=+0.059$).}
\vspace{5pt}
\label{tab:law-3}
\begin{tabular}{l r r r r}
\toprule
 & \multicolumn{2}{c}{mean chars} & \multicolumn{2}{c}{\#placeh.} \\
\cmidrule(lr){2-3} \cmidrule(lr){4-5}
Puzzle & tr & te & tr & te \\
\midrule
Campsite & $+0.094$ & $+0.123$ & $+0.132$ & $+0.145$ \\
CarPainting & $+0.006$ & $+0.015$ & $+0.008$ & $+0.049$ \\
CryptoKKA & $+0.171$ & $+0.132$ & $+0.154$ & $+0.132$ \\
CryptoKPA & $+0.052$ & $+0.028$ & $+0.056$ & $+0.029$ \\
Game24 & $+0.047$ & $+0.025$ & $+0.048$ & $+0.020$ \\
HamiltonianPath & $+0.052$ & $+0.069$ & $+0.011$ & $-0.014$ \\
Kakurasu & $+0.256$ & $+0.119$ & $+0.270$ & $+0.076$ \\
MagicSquare & $+0.069$ & $+0.042$ & $+0.119$ & $+0.061$ \\
StackPermutation & $+0.074$ & $+0.130$ & $+0.102$ & $+0.059$ \\
Sudoku & $+0.069$ & $+0.189$ & $+0.184$ & $+0.200$ \\
Sudoku2 & $+0.102$ & $+0.187$ & $+0.135$ & $+0.135$ \\
SumSkyscraper & $+0.073$ & $+0.016$ & $+0.025$ & $+0.035$ \\
ZebraLogic & $+0.122$ & $+0.118$ & $+0.072$ & $-0.010$ \\
\midrule
$k_{+}/n$ & 11/13 & 6/13 & 9/13 & 7/13 \\
$k_{-}/n$ & 0/13 & 0/13 & 0/13 & 0/13 \\
median $\Delta$ & $+0.073$ & $+0.118$ & $+0.102$ & $+0.059$ \\
\bottomrule
\end{tabular}
\end{table}

\section{Design Space Specifications}
\label{apdx:scope}

We designed the operator set based on an extensive literature review covering contemporary research papers \cite{mass,supernet,hu2025automated,chenAgentVerseFacilitatingMultiAgent2023,qianSCALINGLARGELANGUAGE2025}, agentic system design whitepapers, and guidelines from popular GitHub repositories.
We cover most of the designs from \cite{scalingagent}, with the exception of the Hybrid.
Our set of operators covers commonly used patterns; our aim was to provide the general topologies, while leveraging \aflow{} ability to combine workflows, so that the core topological components could be assembled into complex patterns while making it possible to track operator usage. 

\subsection{Operator Set}
\label{apdx:operator_set}

The operator set is the unit of architectural variation in our search space.
Each operator is a multi-agent \emph{topology primitive} whose call signature exposes a small set of structural knobs (number of agents, number of rounds, prompts) and consumes one or two of the three LLM roles: \emph{worker}, \emph{aggregator} (or its role-equivalents \emph{coordinator}, \emph{moderator}), and \emph{optimizer}.
The six operators currently registered are listed below; Fig.~\ref{fig:operators} draws each of them for $n=3$ workers and $r=2$ rounds together with its call and step counts, and Fig.~\ref{fig:composition} shows how two operators are chained inside \texttt{graph.py}.
All extend a common base class that clamps the multi-agent knobs to safe ranges:  $1 \le \texttt{num\_agents} \le 10$ and $1 \le \texttt{num\_rounds} \le 5$, so that any concrete workflow has a bounded LLM-call budget regardless of the optimizer's choices.
A \emph{round} is one complete pass of the operator's communication pattern (every worker acts once); operators with $\texttt{num\_rounds} > 1$ repeat this pass, each time conditioned on the state produced by the previous pass.
Table~\ref{tab:operators} summarizes the parameters and call counts of each operator.

\begin{table}[t]
\centering
\resizebox{\columnwidth}{!}{%
\begin{tabular}{@{}lllll@{}}
\toprule
\textbf{Operator} & \textbf{Communication} & \textbf{Structure} & \textbf{Params} & \textbf{Calls} \\
\midrule
\textsc{Sequential}          & single agent              & linear pipeline       & $r, \mathbf{p}$           & $|\mathbf{p}|\cdot r$ \\
\textsc{Star}                & isolated workers          & fan-out $\to$ agg     & $n, \mathbf{p}_w, p_a$    & $n + 1$ \\
\textsc{SharedMemory}        & shared doc                & writes $\to$ agg      & $n, r, \mathbf{p}_w, p_a$ & $n\cdot r + 1$ \\
\textsc{Debate}              & all-to-all                & exchange $\to$ agg    & $n, r, \mathbf{p}_w, p_a$ & $n\cdot r + 1$ \\
\textsc{StarW/Coordinator} & coord $\leftrightarrow$ workers & iterative hub-spoke & $n, r, \mathbf{p}_w, p_a$ & $(n+1)\cdot r$ \\
\textsc{DebateW/Moderator} & all-to-all + mod          & debate + mod/round    & $n, r, \mathbf{p}_w, p_a$ & $(n+1)\cdot r$ \\
\bottomrule
\end{tabular}%
}
\caption{\textbf{Composable operator primitives.} $n \in [1,10]$ workers; $r \in [1,5]$ rounds. $\mathbf{p}_w$: list of worker prompts (one per agent); $p_a$: second-role prompt (aggregator / coordinator / moderator). \textsc{Sequential} is single-agent and takes a single prompt list $\mathbf{p}$. See Fig.~\ref{fig:operators} for the topology of each operator and Fig.~\ref{fig:composition} for how operators compose. }
\label{tab:operators}
\end{table}

\paragraph{Sequential}
A pipeline of single-agent calls.
The agent is invoked once per prompt in \texttt{prompts}, with each call receiving the previous call's output as additional context, and the chain is repeated \texttt{num\_rounds} times.
There is no multi-agent coordination: only the worker role is consumed, $\texttt{num\_calls} = \texttt{len(prompts)} \cdot \texttt{num\_rounds}$.
This is the closest the operator set comes to a single-agent baseline.

\paragraph{Star}
Parallel fan-out followed by a single aggregation step.
\texttt{num\_agents} workers run in parallel on the same problem, each with its own prompt; an aggregator then reads the concatenated worker outputs and produces a single answer.
Workers are fully isolated and never see one another's responses.
$\texttt{num\_calls} = \texttt{num\_agents} + 1$, mixing the worker and aggregator roles.

\paragraph{SharedMemory}
Sequential writers to a shared document, with a final aggregation pass.
On each of \texttt{num\_rounds} rounds, the \texttt{num\_agents} workers take turns reading the current document and appending a contribution; the aggregator then produces an answer from the final document.
Communication is indirect through the document state.
$\texttt{num\_calls} = \texttt{num\_agents} \cdot \texttt{num\_rounds} + 1$.

\paragraph{Debate}
All-to-all explicit communication over \texttt{n\_rounds} rounds.
Each round, every agent revises its proposal having seen all other agents' current proposals; a final aggregator call collapses the last round's proposals into a single answer.
Communication is direct, with full proposal visibility at every step.
$\texttt{num\_calls} = \texttt{num\_agents} \cdot \texttt{n\_rounds} + 1$.

\paragraph{StarWithCoordinator}
An iterative hub-and-spoke: the coordinator participates actively across \texttt{n\_rounds} rounds.
Each round, the coordinator broadcasts a directive, the \texttt{num\_agents} workers respond in parallel, and the coordinator reads all responses to issue a revised directive; the directive after the final round is the operator's output.
There is no worker-to-worker communication; the aggregator role is repurposed as the coordinator.
$\texttt{num\_calls} = (\texttt{num\_agents} + 1) \cdot \texttt{n\_rounds}$.

\paragraph{DebateWithModerator}
A moderated variant of \textsc{Debate}: a moderator agent injects a summary or steering prompt after every all-to-all exchange.
Each agent now sees both its peers' proposals and the moderator's most recent intervention before producing its next response, and the moderator's output on the final round is the operator's answer.
The aggregator role is repurposed as the moderator.
$\texttt{num\_calls} = (\texttt{num\_agents} + 1) \cdot \texttt{n\_rounds}$, but with a richer signal at each step than \textsc{StarWithCoordinator}.

\subsection{Parent Selection Policy}
\label{app:sampling}
 
Parent workflow selection at each \aflow{} expansion step follows the mixed policy of \citet{aflow}. Let $s_1, \dots, s_n$ denote the scores of the top-$k$ previously explored candidates ($k = 4$), with $s_{\max} = \max_i s_i$. The probability of selecting candidate $i$ is
\begin{equation}
\resizebox{0.88\columnwidth}{!}{$\displaystyle P_{\mathrm{mixed}}(i) \;=\; \lambda \cdot \frac{1}{n} \;+\; (1 - \lambda) \cdot \frac{\exp\bigl(\alpha \cdot (s_i - s_{\max})\bigr)}{\sum_{j=1}^{n} \exp\bigl(\alpha \cdot (s_j - s_{\max})\bigr)},$}
\label{eq:mixture}
\end{equation}
where $\lambda = 0.2$ controls the trade-off between uniform sampling and score-weighted sampling, and $\alpha = 0.4$ controls the temperature of the score-weighted component. The shift by $s_{\max}$ stabilizes the softmax against numerical overflow without altering the resulting distribution. The uniform component prevents collapse onto the optimizer's prior over high-frequency operators (cf.\ Table~\ref{tab:operator_percentages}) and ensures broad trajectory coverage across rounds (\S~\ref{app:theory}).
 
\subsection{\aflow{} Integration}
\label{apdx:aflow}
 
The \aflow{} codebase \citep{aflow} was modified along three axes. First, the operator registry was replaced with the six primitives of \S\ref{apdx:operator_set}, with corresponding updates to the optimizer-facing operator manual injected into the modification prompt (\S\ref{app:prompts}). Second, the dataset adapter was extended to interface with the \enigmata{} generators and verifiers; this required matching \enigmata{}'s per-puzzle scoring contracts to \aflow{}'s evaluator interface and routing the difficulty-bundle parameter through the existing task-config plumbing. Third, the per-round telemetry was extended to record 1) per-call worker/aggregator role attribution, and 2) the prompts emitted by the optimizer for each instantiation.

\subsection{\enigmata{} Specification}
\label{app:enigmata}
 
\enigmata{} \citep{enigmata} bundles 36 abstract-reasoning puzzles, each paired with a parametric generator and a deterministic, rule-based verifier. The verifier performs an exact match against canonicalized ground-truth output and admits no partial credit, with the per-instance score $\in \{0, 1\}$. We restrict our experiments to the $30$ puzzles for which both generator and verifier expose programmatic interfaces; the remaining six puzzles are excluded.
Difficulty bundles are inherited from the \enigmata{} partition, which was calibrated against GPT-4o pass rates \citep{enigmata}: $27$ of the $30$ retained puzzles expose the full $\{\text{Easy}, \text{Medium}, \text{Hard}\}$ bundle, while three (Hitori, LightUp, and TicTacToe) expose only $\{\text{Medium}, \text{Hard}\}$.
 
For each (puzzle, bundle) cell, we sample $M = 20$ instances at search initialization and reuse this fixed set across all $K = 5$ \aflow{} searches and all workflow rounds within a search. Reusing instances within a cell ensures that score variation across workflows reflects workflow structure rather than instance resampling. Mixed-difficulty bundles are left to future work. The category breakdown of the different puzzles is reported in Table~\ref{tab:baselines}.
 
\subsection{Filtered Puzzles and Failure Modes}
\label{app:failures}
 
Of the $30$ retained puzzles, four (EightPuzzle, NinePuzzle, FifteenPuzzle, and SixteenPuzzle), all in the Sequential category, were excluded from the analysis corpus. On these puzzles, the maximum score across the entire set was $0$: no workflow ever produced a verifier-accepted solution at any difficulty level. Inspection of representative completions confirmed that the failure mode is uniform across operators and difficulty bundles, with the worker model either emitting an incorrect move sequence or asserting that no solution exists. Because admission to the predictability gate requires non-degenerate score variance within a puzzle, these four puzzles cannot contribute to any per-puzzle effect-size estimate; retaining them would inflate $n_{\mathrm{total}}$ without adding statistical power. The filtered analysis corpus therefore comprises $26$ puzzles and $28{,}568$ workflows (Table~\ref{tab:dataset_structure}).
 
A secondary failure mode, applying to the executable-workflow filter rather than to whole puzzles, removed $1{,}031$ workflows ($3.0\%$ of the post-bundle corpus) for which the \aflow{} Optimizer produced a graph that could not be evaluated, typically because a custom prompt referenced an undefined placeholder or a deleted prompt key. 
These failures are uniformly distributed across the (puzzle, bundle, model) grid and do not bias the structural feature distribution conditional on successful execution.

\section{Analysis Phase Methodology}
\label{apdx:methodology}
\begin{figure}[htbp]
    \centering
    \includegraphics[width=\linewidth]{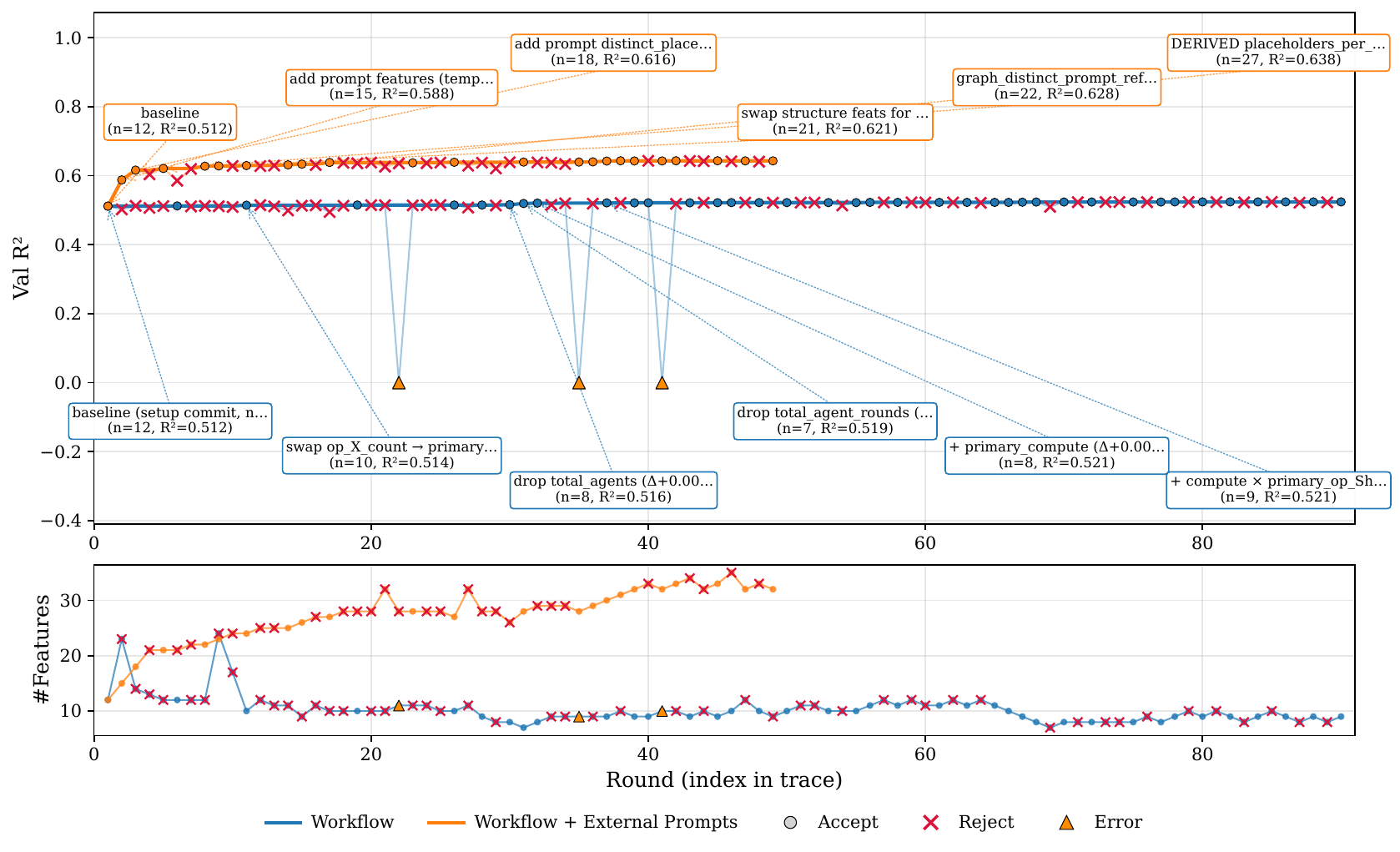}
    \caption{Example of two \autoresearch{} searches, one exploring with access to only the executable workflow code without custom prompts, the other with access to both. The number of rounds differs based on when the \autoresearch{} instance considers it is stalling.}
    \label{fig:autoresearch_trace}
\end{figure}

\subsection{Feature Engineering and Models}
\label{apdx:feature_dictionary}

The feature set $\ba = \phi(W)$ is computed only from the workflow's own source -- the operator graph and the prompts -- plus the two allow-listed problem-side descriptors (puzzle family and difficulty bundle).
The shared operator library, while readable as a reference, is identical across every workflow and therefore cannot produce a feature.
The base parsers cover counts and structural summaries of operator calls, prompt-length statistics, placeholder topology, and a handful of style markers (step-by-step, format blocks); the autonomous loop iterates on them.

\paragraph{Train / val / test split provenance}
\label{apdx:split}
The three-way split is assigned at the level of the \aflow{} \emph{search}, not the workflow.
Each search is identified by a stable hash of its full configuration; the assignment is deterministic per hash with a fixed seed (so a re-run on the same configuration lands in the same split), and the corpus-flattening step only inherits this assignment without resampling.
After flattening, every workflow lives at a content-free identifier $\texttt{workflow\_id} = \text{sha1}(\texttt{search\_hash}\!:\!\texttt{round})[:16]$; the legacy \texttt{(task, round, search\_hash)} path components are removed from the directory structure, and the two task-/round-dependent imports inside each workflow's graph source are rewritten to neutral, content-free forms.
Together, these guarantee that the workflow source is independent of its on-disk path.
The reproducibility contract on the loop is stronger than just split determinism: on any commit, re-running the analysis script after the flattening step must reproduce the committed findings snapshot and the structured metrics record byte-for-byte.
This is enforced through pinned random seeds inside every random-state-bearing step of the regression pipeline, a frozen dependency environment, and a prohibition on machine-local state beyond the workflow corpus and a single targets cache.

\paragraph{Identifier exclusions}
The row-level identifiers (\texttt{workflow\_id}, round index, configuration hash) are forbidden as features, including in any reconstructed or hashed form. The split column itself is also forbidden.
 
\paragraph{Target-derived feature exclusions}
No feature may depend on the primary target $m$ or on the workflow's total cost on any row. This prohibition extends to target encoding of categoricals (smoothed or otherwise), per-group target aggregates, residuals, ranks, and label-derived difficulty priors. A label-derived feature is, by construction, never a property of the workflow's design.

\subsection{\autoresearch{} Details}
\label{apdx:loop_log}

Every iteration of the autonomous loop atomically commits, alongside the analysis script, a fixed set of artifacts that together render any past iteration replayable from the git history alone:
\begin{itemize}
\item \texttt{metrics.json}: structured metrics, feature names, and top coefficients (the time-series source for Fig.~\ref{fig:autoresearch_trace});
\item \texttt{feature\_names.json}: the active feature list as a flat array;
\item \texttt{train.py}: the training script driving feature engineering and surrogate fitting.
\end{itemize}
The fitted model and preprocessor binaries are written outside the git tree and overwritten on each run; checking out a past commit and re-running the analysis script regenerates them, and serves as the byte-for-byte reproducibility check.

\paragraph{Per-commit log}
A per-commit log table accompanies the artifacts and functions as the running lab notebook of the autonomous loop. Every attempt, whether accepted, discarded, or aborted by a runtime error, appears in this log because the branch is append-only by construction. The verdict labels are the agent's own classification, recorded at the time of commit; they are included as a reproducibility artifact rather than a primary result, and are not used as features by the surrogate.

\subsection{Caveats and Mitigated Failure Modes}
\label{app:caveats}
 
The combination of an LLM-driven feature proposer and a sandboxed evaluation loop is, by construction, susceptible to feature constructions that exploit incidental dataset structure rather than workflow design. Two failure modes were observed during early \autoresearch{} runs and addressed before the design principles reported were finalized.
 
\paragraph{Round index as a feature}
An early proposal used the \aflow{} round index as a direct numerical feature, exploiting the fact that later-round workflows tend to be selected from higher-scoring parents and therefore inherit a positive score correlation that is unrelated to architectural design. The mitigation strips the round index from the feature-side metadata exposed to the feature-mining loop and restricts the available artifacts to the Python implementations of the workflows. The round index is retained in evaluation metadata for analysis bookkeeping but is never visible to $\phi$.
 
\paragraph{Train-mean leakage}
A second early proposal computed the mean training-set score per (puzzle, bundle) cell and added it as a feature on the validation and test rows, yielding spuriously high $R^2$. The mitigation extends the sandbox prompt with an explicit prohibition on aggregate statistics derived from the target column, regardless of split.

\subsection{Prompts} \label{app:prompts}
We provide LLM-MAS adapted parts of the prompts used by \aflow{} to conduct the workflow searches, tailored to our operator set.
\begin{promptbox}[title={\aflow{} MAS Optimize Prompt}, fontupper=\scriptsize, left=4pt, right=4pt]
\begin{Verbatim}[fontsize=\scriptsize,breaklines=true,breakanywhere=true,breaksymbolleft={}]
You are building a Graph and corresponding Prompt to jointly solve
{type} problems using multi-agent system (MAS) operators.
Referring to the given graph and prompt, which forms a basic example
of a {type} solution approach,
please reconstruct and optimize them.
You can add, modify, or delete nodes, parameters, or prompts. 
Include your
single modification in XML tags in your reply. 
Ensure they are complete and correct to avoid runtime failures.
The available operators are multi-agent topologies:
Sequential, Star, SharedMemory, Debate, StarWithCoordinator, 
DebateWithModerator.
Each operator takes self.llm (worker) and self.aggregator_llm 
(aggregator/coordinator/moderator) as LLMs
— do NOT create or reference other LLMs.
You can combine operators, chain them, or adjust their parameters 
(num_agents, n_rounds, prompts) to improve performance.
The graph complexity should not exceed 10 operator invocations.
Use logical and control flow (IF-ELSE, loops) for enhanced graph design.

Ensure that all prompts required by the current graph from prompt_custom
are included. 
Exclude any unused prompts.
Output the modified graph and all necessary prompts in
prompt_custom (if needed).
The generated prompt must not contain any placeholders other than the 
ones supported by the operators.
Considering information loss, complex graphs may yield better results, 
but insufficient information transmission can omit the solution.
It's crucial to include necessary context during the process.
\end{Verbatim}
\end{promptbox}

\begin{promptbox}[title={\aflow{} MAS Component Usage}, fontupper=\scriptsize, left=4pt, right=4pt]
Here are examples of using the MAS operators in the graph: \\

\begin{Verbatim}[fontsize=\scriptsize,breaklines=true,breakanywhere=true,breaksymbolleft={}]
# Sequential: linear pipeline, each step gets the previous output
seq = operator.Sequential(llm=self.llm, num_rounds=2, 
    prompts=["Solve: {problem}",
    "Review and improve: {previous_output}"])
result = await seq(problem)

# Star: parallel workers then aggregation
star = operator.Star(worker_llm=self.llm,
aggregator_llm=self.aggregator_ll
, num_agents=3,
    worker_prompts=["Solve this: {problem}"] * 3,
    aggregator_prompt="Given these solutions: {worker_outputs}
    Pick the best for: {problem}")
result = await star(problem)

# Debate: agents debate then aggregation
debate = operator.Debate(worker_llm=self.llm, 
    aggregator_llm=self.aggregator_llm,
    num_agents=3,
    worker_prompts=["Solve: {problem} Other proposals:
    {all_proposals}"] * 3,
    aggregator_prompt="Final answer for {problem} given:
    {all_proposals}",
    n_rounds=2)
result = await debate(problem)

# All MAS operators return dict with key 'response'
answer = result['response']

`
The available LLMs are self.llm (worker) and self.aggregator_llm 
(aggregator/coordinator/moderator).

Supported placeholders per operator:
- Sequential: {problem}, {previous_output}
- Star: worker_prompts use {problem}; aggregator_prompt uses {problem},
{worker_outputs}
- SharedMemory: worker_prompts use {problem}, {document}, {agent_id}, 
    {round};
    aggregator_prompt uses {problem}, {document}
- Debate: worker_prompts use {problem}, {all_proposals}, {round},
    {agent_id}; aggregator_prompt uses {problem}, {all_proposals}
- StarWithCoordinator: worker_prompts use {problem}, {directive},
    {round}, {agent_id}; coordinator_prompt uses {problem},
    {worker_outputs}, {round}, {is_final}
- DebateWithModerator: agent_prompts use {problem}, {all_proposals},
    {moderator_input}, {round}, {agent_id};
    moderator_prompt uses {problem}, {all_proposals}, {round},
    {is_final}

You can also define custom prompts in prompt_custom and reference them.

**Combining multiple operators or chaining them can enhance performance.
Try different topologies.**
\end{Verbatim}
\end{promptbox}

\section{Corpus Details and Additional Analyses}
\label{apdx:add_res}

\subsection{Statistics}

\begin{table}[htbp]
    \centering
    \begin{tabular}{ccc}
        \toprule
         Split & Count & Percentage \\
         \midrule
         Total & 28,568 & 100.0\% \\
         \midrule
         Train & 19,822 & 69.4\% \\
         Validation & 4,278 & 15.0\% \\
         Test & 4,468 & 15.6\% \\
         \bottomrule
    \end{tabular}
    \vspace{5pt}
    \caption{Effective workflow counts across splits. Of 1,701 in-scope completed AFlow searches, each generating up to 20 workflows (34,020 candidates), 32,989 are executable. A further drop to 28,568 workflows removes the four filtered sliding-tile puzzles (Eight/Fifteen/Nine/SixteenPuzzle), where no workflow ever scored above zero. These are then divided into the train/val/test split.}
    \label{tab:dataset_structure}
\end{table}

\subsection{Cluster-Robust Inference and Small-Cluster Coverage}
\label{app:cluster-robust}
 
The per-cell confidence intervals reported in \S\ref{subsec:analysis} and in Tables~\ref{tab:law-2}--\ref{tab:law-3} are computed using cluster-Wald standard errors with the \aflow{} search identifier as the clustering unit.
Clustering at the search level rather than at the workflow level is the appropriate choice: workflows generated within the same MCTS trace are not independent draws from the workflow population, since the optimizer conditions each expansion on the score history of its parent.
The cluster-Wald estimator is consistent in the large-cluster-count limit, with standard recommendations requiring at least $G \geq 15$ clusters per arm for nominal coverage.
 
\paragraph{Train-split cluster support}
On the training split, per-arm cluster counts across the cells in Tables~\ref{tab:law-2}--\ref{tab:law-3} range from $19$ to $47$. A single contrast (TicTacToe~$\times$~\texttt{op\_SharedMemory\_count}) sits at the lower bound with $G = 19$ clusters in its smaller arm; all other train-side contrasts have $G \geq 24$ per arm. The train-split CIs therefore lie within the regime in which cluster-Wald inference is well-supported, and the train-side strict-sign verdicts ($k_+ / k_-$) are unaffected by the issue described below.
 
\paragraph{Test-split cluster support}
The held-out test split is, by construction, more sparsely supported at the cluster level: the search-level split assigns each \aflow{} search to exactly one of train, validation, and test, and the test split contains $\sim 15\%$ of searches. Across the cells reported in the test columns of Tables~\ref{tab:law-2}--\ref{tab:law-3}, per-arm cluster counts range from $4$ to $14$, with the majority of cells below $G = 10$. The least-supported cells, both with $G = 4$ per arm, would fall outside the asymptotic regime.

\begin{table}[htbp]
    \centering
    \resizebox{\columnwidth}{!}{%
    \begin{tabular}{l|c|c|c|c}
        \toprule
        Task & Easy (\$) & Medium (\$) & Hard (\$) & Mean (\$) \\
        \midrule
        StarBattle & \$0.0110 & \$0.0096 & \$0.0099 & \$0.0102 \\
        SumSkyscraper & \$0.0108 & \$0.0092 & \$0.0102 & \$0.0100 \\
        Sudoku & \$0.0096 & \$0.0105 & \$0.0090 & \$0.0097 \\
        ZebraLogic & \$0.0062 & \$0.0097 & \$0.0116 & \$0.0091 \\
        Campsite & \$0.0065 & \$0.0104 & \$0.0102 & \$0.0091 \\
        Kakurasu & \$0.0094 & \$0.0063 & \$0.0113 & \$0.0090 \\
        Hitori & -- & \$0.0093 & \$0.0086 & \$0.0090 \\
        HamiltonianCycle & \$0.0083 & \$0.0073 & \$0.0087 & \$0.0081 \\
        LightUp & -- & \$0.0070 & \$0.0082 & \$0.0076 \\
        CarPainting & \$0.0079 & \$0.0070 & \$0.0075 & \$0.0075 \\
        Skyscraper & \$0.0067 & \$0.0074 & \$0.0076 & \$0.0072 \\
        HamiltonianPath & \$0.0058 & \$0.0072 & \$0.0085 & \$0.0072 \\
        Slant & \$0.0069 & \$0.0071 & \$0.0065 & \$0.0068 \\
        Binario & \$0.0062 & \$0.0069 & \$0.0064 & \$0.0065 \\
        MagicSquare & \$0.0054 & \$0.0059 & \$0.0077 & \$0.0063 \\
        StackPermutation & \$0.0054 & \$0.0067 & \$0.0063 & \$0.0061 \\
        Minesweeper & \$0.0059 & \$0.0058 & \$0.0067 & \$0.0061 \\
        Maze & \$0.0064 & \$0.0058 & \$0.0049 & \$0.0057 \\
        Twiddle & \$0.0040 & \$0.0072 & \$0.0056 & \$0.0056 \\
        Countdown & \$0.0057 & \$0.0061 & \$0.0047 & \$0.0055 \\
        SixteenPuzzle & \$0.0047 & \$0.0043 & \$0.0049 & \$0.0046 \\
        FifteenPuzzle & \$0.0044 & \$0.0044 & \$0.0045 & \$0.0044 \\
        Sudoku2 & \$0.0045 & \$0.0044 & \$0.0044 & \$0.0044 \\
        NaturalLanguageNavigation & \$0.0030 & \$0.0041 & \$0.0060 & \$0.0043 \\
        Game24 & \$0.0041 & \$0.0045 & \$0.0043 & \$0.0043 \\
        EightPuzzle & \$0.0042 & \$0.0039 & \$0.0043 & \$0.0041 \\
        NinePuzzle & \$0.0040 & \$0.0041 & \$0.0037 & \$0.0040 \\
        TicTacToe & -- & \$0.0037 & \$0.0041 & \$0.0039 \\
        CryptoKKA & \$0.0040 & \$0.0029 & \$0.0030 & \$0.0033 \\
        CryptoKPA & \$0.0037 & \$0.0020 & \$0.0031 & \$0.0030 \\
        \midrule
        \textit{Avg.} & \$0.0061 & \$0.0064 & \$0.0067 & \$0.0064 \\
        \bottomrule
    \end{tabular}%
    }
    \vspace{5pt}
    \caption{Average cost in USD to evaluate one workflow against a single target puzzle instance, by task and difficulty. ``--'' marks (puzzle, difficulty) cells that Enigmata's generator does not provide.}
    \label{tab:cost_per_instance}
\end{table}

\begin{table}[htbp]
    \centering
    \resizebox{\columnwidth}{!}{%
    \begin{tabular}{lrrcccc}
        \toprule
        \multirow{2}{*}{Operator} & \multirow{2}{*}{\# Workflows} & \multirow{2}{*}{Percentage} & \multicolumn{2}{c}{Score} & \multicolumn{2}{c}{Cost} \\
        \cmidrule(lr){4-5} \cmidrule(lr){6-7}
         & & & w & w/o & w & w/o \\
        \midrule
        Sequential & 13,473 & 47.16 & 0.14 & 0.19 & 0.14 & 0.15 \\
        Star & 12,716 & 44.51 & 0.16 & 0.17 & 0.13 & 0.16 \\
        Debate & 6,373 & 22.31 & 0.18 & 0.17 & 0.13 & 0.15 \\
        SharedMemory & 5,948 & 20.82 & 0.23 & 0.15 & 0.25 & 0.12 \\
        DebateWithModerator & 3,382 & 11.84 & 0.19 & 0.16 & 0.19 & 0.14 \\
        StarWithCoordinator & 2,502 & 8.76 & 0.16 & 0.17 & 0.24 & 0.14 \\
        \bottomrule
    \end{tabular}%
    }
    \vspace{5pt}
    \caption{Operator presence over the filtered analysis corpus (28{,}568 workflows). For each operator we contrast the mean of \emph{score} and \emph{cost} between workflows that contain the operator (\emph{w}) and those that don't (\emph{w/o}).}
    \label{tab:operator_percentages}
\end{table}

\begin{table}[htbp]
    \centering
    \resizebox{\columnwidth}{!}{%
    \begin{tabular}{lcccccc}
        \toprule
        \multirow{2}{*}{Operator combination} & \multirow{2}{*}{\# WFs} & \multirow{2}{*}{Percentage} & \multicolumn{2}{c}{Score} & \multicolumn{2}{c}{Cost} \\
        \cmidrule(lr){4-5} \cmidrule(lr){6-7}
         & & & w & w/o & w & w/o \\
        \midrule
        Star $\to$ Seq & 4,564 & 15.98 & 0.14 & 0.17 & 0.12 & 0.15 \\
        Star & 4,125 & 14.44 & 0.15 & 0.17 & 0.07 & 0.16 \\
        Debate & 2,346 & 8.21 & 0.17 & 0.17 & 0.09 & 0.15 \\
        SharedMemory $\to$ Seq & 2,158 & 7.55 & 0.18 & 0.17 & 0.23 & 0.14 \\
        Seq & 2,078 & 7.27 & 0.09 & 0.17 & 0.05 & 0.16 \\
        Debate $\to$ Seq & 2,041 & 7.14 & 0.14 & 0.17 & 0.12 & 0.15 \\
        SharedMemory & 1,748 & 6.12 & 0.25 & 0.16 & 0.20 & 0.14 \\
        DebateWithModerator & 1,510 & 5.29 & 0.19 & 0.17 & 0.12 & 0.15 \\
        StarWithCoordinator & 1,085 & 3.80 & 0.15 & 0.17 & 0.21 & 0.15 \\
        StarWithCoordinator $\to$ Seq & 741 & 2.59 & 0.13 & 0.17 & 0.24 & 0.14 \\
        \bottomrule
    \end{tabular}%
    }
    \vspace{5pt}
    \caption{Top-10 most-used operator combinations over the filtered analysis corpus (28{,}568 workflows), ranked by frequency. Operators in a sequence are connected by $\to$. For each row, we contrast the mean of \emph{score} and \emph{cost} between workflows that use the exact sequence (\emph{w}) and the rest of the corpus (\emph{w/o}).}
    \label{tab:most_common_operator_combinations}
\end{table}

\begin{table*}[t]
    \centering
    \footnotesize
    \begin{tabular}{lcccccccccc}
        \toprule
        Puzzle & \# WFs & Mean & Std & Min & 25\% & 50\% & 75\% & Max & Skew & Kurt. \\
        \midrule
        \multicolumn{11}{l}{\textit{Graph}} \\
        HamilCycle & 1,164 & 0.073 & 0.101 & 0.00 & 0.00 & 0.00 & 0.10 & 0.40 & 1.40 & 1.10 \\
        HamilPath & 1,147 & 0.150 & 0.158 & 0.00 & 0.00 & 0.10 & 0.25 & 0.70 & 0.96 & -0.05 \\
        Maze & 1,075 & 0.479 & 0.221 & 0.00 & 0.35 & 0.50 & 0.65 & 1.00 & -0.21 & -0.29 \\
        NLNavigate & 1,120 & 0.561 & 0.322 & 0.00 & 0.30 & 0.65 & 0.85 & 1.00 & -0.36 & -1.18 \\
        \midrule
        \multicolumn{11}{l}{\textit{Logic}} \\
        ZebraLogic & 1,159 & 0.184 & 0.266 & 0.00 & 0.00 & 0.05 & 0.30 & 1.00 & 1.42 & 0.90 \\
        \midrule
        \multicolumn{11}{l}{\textit{Grid}} \\
        Binario & 1,120 & 0.208 & 0.249 & 0.00 & 0.00 & 0.10 & 0.30 & 1.00 & 1.44 & 1.50 \\
        Campsite & 1,136 & 0.088 & 0.167 & 0.00 & 0.00 & 0.00 & 0.10 & 0.85 & 2.12 & 3.81 \\
        MagicSquare & 1,166 & 0.131 & 0.175 & 0.00 & 0.00 & 0.05 & 0.15 & 0.65 & 1.48 & 1.02 \\
        Skyscraper & 1,182 & 0.003 & 0.014 & 0.00 & 0.00 & 0.00 & 0.00 & 0.15 & 4.76 & 26.62 \\
        StarBattle & 1,197 & 0.024 & 0.067 & 0.00 & 0.00 & 0.00 & 0.00 & 0.50 & 3.59 & 13.74 \\
        Sudoku & 1,199 & 0.300 & 0.355 & 0.00 & 0.00 & 0.10 & 0.55 & 1.00 & 0.83 & -0.81 \\
        Sudoku2 & 1,178 & 0.698 & 0.374 & 0.00 & 0.55 & 0.90 & 1.00 & 1.00 & -1.04 & -0.52 \\
        SumSkyscrap & 1,152 & 0.050 & 0.140 & 0.00 & 0.00 & 0.00 & 0.00 & 0.90 & 3.22 & 10.01 \\
        \midrule
        \multicolumn{11}{l}{\textit{Crypto}} \\
        CryptoKKA & 1,114 & 0.121 & 0.210 & 0.00 & 0.00 & 0.00 & 0.15 & 1.00 & 2.08 & 3.67 \\
        CryptoKPA & 1,191 & 0.033 & 0.077 & 0.00 & 0.00 & 0.00 & 0.05 & 0.65 & 3.74 & 17.38 \\
        \midrule
        \multicolumn{11}{l}{\textit{Arithmetic}} \\
        Countdown & 1,137 & 0.038 & 0.089 & 0.00 & 0.00 & 0.00 & 0.05 & 0.70 & 3.66 & 15.95 \\
        Game24 & 1,144 & 0.107 & 0.122 & 0.00 & 0.00 & 0.05 & 0.15 & 0.70 & 1.32 & 1.17 \\
        \midrule
        \multicolumn{11}{l}{\textit{Search}} \\
        Hitori & 768 & 0.039 & 0.102 & 0.00 & 0.00 & 0.00 & 0.05 & 0.85 & 4.49 & 24.24 \\
        Kakurasu & 1,159 & 0.136 & 0.240 & 0.00 & 0.00 & 0.00 & 0.15 & 0.95 & 1.81 & 2.13 \\
        LightUp & 732 & 0.000 & 0.005 & 0.00 & 0.00 & 0.00 & 0.00 & 0.05 & 10.93 & 117.82 \\
        Minesweeper & 1,079 & 0.006 & 0.038 & 0.00 & 0.00 & 0.00 & 0.00 & 0.60 & 10.32 & 129.82 \\
        Slant & 1,136 & 0.000 & 0.002 & 0.00 & 0.00 & 0.00 & 0.00 & 0.05 & 23.80 & 565.49 \\
        TicTacToe & 737 & 0.376 & 0.203 & 0.00 & 0.25 & 0.40 & 0.50 & 0.85 & -0.27 & -0.63 \\
        \midrule
        \multicolumn{11}{l}{\textit{Sequential}} \\
        CarPainting & 1,159 & 0.044 & 0.077 & 0.00 & 0.00 & 0.00 & 0.05 & 0.40 & 2.00 & 3.61 \\
        EightPuzzle & 1,129 & 0.000 & 0.000 & 0.00 & 0.00 & 0.00 & 0.00 & 0.00 & 0.00 & 0.00 \\
        FifteenPuzzle & 1,152 & 0.000 & 0.000 & 0.00 & 0.00 & 0.00 & 0.00 & 0.00 & 0.00 & 0.00 \\
        NinePuzzle & 1,058 & 0.000 & 0.000 & 0.00 & 0.00 & 0.00 & 0.00 & 0.00 & 0.00 & 0.00 \\
        SixteenPuzzle & 1,082 & 0.000 & 0.000 & 0.00 & 0.00 & 0.00 & 0.00 & 0.00 & 0.00 & 0.00 \\
        StackPerm & 1,077 & 0.508 & 0.253 & 0.00 & 0.35 & 0.45 & 0.65 & 1.00 & 0.26 & -0.43 \\
        Twiddle & 1,140 & 0.014 & 0.052 & 0.00 & 0.00 & 0.00 & 0.00 & 0.45 & 4.61 & 23.52 \\
        \bottomrule
    \end{tabular}
    \vspace{5pt}
    \caption{Per-puzzle distribution of \texttt{score} over the full corpus of 32{,}989 workflows on the 30 Enigmata puzzles. Rows are grouped by category (Graph $\to$ Logic $\to$ Grid $\to$ Crypto $\to$ Arithmetic $\to$ Search $\to$ Sequential), alphabetical within category. The four sliding-tile puzzles in the \emph{Sequential} category (Eight/Fifteen/Nine/SixteenPuzzle) have all-zero distributions; they are excluded from the analysis corpus used by the other tables.}
    \label{tab:enigmata_score_distribution}
\end{table*}

\begin{table*}[t]
\footnotesize
\centering
\begin{tabular}{l l c c c c c c}
\toprule
Category & Puzzle & GPT-5.4 & GPT-5.4-mini & GPT-5.4-nano & $W_{\mu}$ &  $W_{max}$ & USD(\$)@max \\
\midrule
\multirow{2}{*}{Arithmetic} & Countdown & 0.033 & 0.017 & 0.017 & 0.173 & 0.700 & 0.228 \\
 & Game24 & 0.150 & 0.083 & 0.050 & 0.288 & 0.700 & 0.220 \\
\midrule
\multirow{2}{*}{Crypto} & CryptoKKA & 0.267 & 0.067 & 0.000 & 0.412 & 1.000 & 0.081 \\
 & CryptoKPA & 0.050 & 0.050 & 0.000 & 0.139 & 0.650 & 0.238 \\
\midrule
\multirow{4}{*}{Graph} & HamilCycle & 0.133 & 0.017 & 0.000 & 0.228 & 0.400 & 0.080 \\
 & HamilPath & 0.300 & 0.100 & 0.017 & 0.363 & 0.700 & 0.233 \\
 & Maze & 0.717 & 0.533 & 0.167 & 0.759 & 1.000 & 0.123 \\
 & NLNavigate & 0.100 & 0.150 & 0.083 & 0.891 & 1.000 & 0.024 \\
\midrule
\multirow{8}{*}{Grid} & Binario & 0.150 & 0.217 & 0.133 & 0.549 & 1.000 & 0.123 \\
 & Campsite & 0.450 & 0.100 & 0.000 & 0.381 & 0.850 & 0.243 \\
 & MagicSquare & 0.033 & 0.000 & 0.000 & 0.359 & 0.650 & 0.106 \\
 & Skyscraper & 0.050 & 0.017 & 0.017 & 0.016 & 0.150 & 0.475 \\
 & StarBattle & 0.033 & 0.000 & 0.000 & 0.121 & 0.500 & 0.501 \\
 & Sudoku & 0.583 & 0.550 & 0.300 & 0.443 & 1.000 & 0.120 \\
 & Sudoku2 & 0.983 & 0.917 & 0.717 & 0.954 & 1.000 & 0.018 \\
 & SumSkyscrap & 0.117 & 0.017 & 0.000 & 0.147 & 0.900 & 0.312 \\
\midrule
\multirow{1}{*}{Logic} & ZebraLogic & 0.517 & 0.217 & 0.017 & 0.377 & 1.000 & 0.200 \\
\midrule
\multirow{6}{*}{Search} & Hitori & 0.025 & 0.000 & 0.000 & 0.202 & 0.850 & 0.362 \\
 & Kakurasu & 0.217 & 0.033 & 0.000 & 0.491 & 0.950 & 0.195 \\
 & LightUp & 0.000 & 0.000 & 0.000 & 0.004 & 0.050 & 0.061 \\
 & Minesweeper & 0.000 & 0.000 & 0.000 & 0.043 & 0.600 & 0.349 \\
 & Slant & 0.000 & 0.000 & 0.000 & 0.002 & 0.050 & 0.315 \\
 & TicTacToe & 0.675 & 0.125 & 0.225 & 0.641 & 0.850 & 0.066 \\
\midrule
\multirow{7}{*}{Sequential} & CarPainting & 0.083 & 0.150 & 0.017 & 0.124 & 0.400 & 0.473 \\
 & EightPuzzle & 0.000 & 0.000 & 0.000 & 0.000 & 0.000 & 0.014 \\
 & FifteenPuzzle & 0.000 & 0.000 & 0.000 & 0.000 & 0.000 & 0.015 \\
 & NinePuzzle & 0.000 & 0.000 & 0.000 & 0.000 & 0.000 & 0.013 \\
 & SixteenPuzzle & 0.000 & 0.000 & 0.000 & 0.000 & 0.000 & 0.015 \\
 & StackPerm & 0.467 & 0.433 & 0.150 & 0.849 & 1.000 & 0.062 \\
 & Twiddle & 0.200 & 0.033 & 0.017 & 0.061 & 0.450 & 0.075 \\
\bottomrule
\end{tabular}
\vspace{5pt}
\caption{Single-backbone baselines: One-shot pass rate of GPT-5.4, GPT-5.4-mini, and GPT-5.4-nano evaluated directly on the per-puzzle bundle without scaffolding, alongside workflow-level statistics from our workflow corpus: mean score $W_{\mu}$, maximum score $W_{\max}$, and the per-instance inference cost of the maximum-scoring workflow ($\$@\max$). Puzzles are grouped by category. The four all-zero sliding-tile puzzles are listed for completeness and excluded from the analysis corpus.}
\label{tab:baselines}
\end{table*}

\subsection{Candidate Hypothesis Set}
\label{app:add_laws}

\begin{table}[t]
\centering
\resizebox{\columnwidth}{!}{%
\begin{tabular}{@{}llcl@{}}
\toprule
 & \textbf{Granularity} & \textbf{Prior Rule?} & \textbf{Representative Mined Principle} \\
\midrule
\textbf{RQ1} & \textit{Entry-Point Topology}     & \cmark & \textsc{Star} as primary operator lowers score \\
\textbf{RQ2} & \textit{Communication State}      & \xmark & \textsc{SharedMemory} outperforms dialogue \\
\textbf{RQ3} & \textit{Sub-Topological Prompts}  & \xmark & Prompt specificity (length, \texttt{\#placeh.}) helps \\
\bottomrule
\end{tabular}%
}
\vspace{5pt}
\caption{\textbf{Three Granularities of Mined Design Principles.} We evaluate our framework across progressively finer levels of workflow specification: from verifying an established macro-topological failure mode (\textbf{RQ1}), to comparing blackboard-style shared state against conversational topologies (\textbf{RQ2}), to mining sub-topological code and prompt template structure (\textbf{RQ3}).}
\label{tab:validation-ladder}
\end{table}

We explored various styles of prompts and constraints for the \autoresearch{} loop to mitigate overfitting, leading to different exploration trajectories and saving protocols; we narrowed our focus to $300$ feature-engineering trials spread across $10$ main \autoresearch{} searches. We used features from these different search traces to generate $70$ candidate binarized feature contrasts (spanning the three granularities in Table~\ref{tab:validation-ladder}), covering hypotheses related to prompt information, number of rounds and agents in workflows, operator positions, and derived features combining merging features.

For instance the two search traces from Fig.~\ref{fig:autoresearch_trace} cover two setups, one with access to both \texttt{graph.py} (containing the executable workflow code) and \texttt{prompt.py} (custom generated prompts), and the other with access only to \texttt{graph.py}. 
The search with access to prompt data ran for $49$ trials, $46.9\%$ of which were accepted and grew the feature set from 12 to 32. The search with access only to \texttt{graph.py} started with an initial set of 12 features and decreased it to 9 features over 90 trials.
Not all trials necessarily optimize features, as the automated search sometimes attempts to optimize model hyperparameters or change the model class directly.

\section{Single-Backbone Baseline Performance}
\label{app:baseline}
To contextualize the workflow-level scores, Table~\ref{tab:baselines} reports the one-shot pass rate of the three model backbones (GPT-5.4, GPT-5.4-mini, GPT-5.4-nano) used as the worker and aggregator policies in our experiments, evaluated directly on the per-puzzle bundle without any multi-agent scaffolding. The same table reports the workflow-level mean score $\bar{\mu}_W$, the workflow-level maximum score $\bar{\mu}_W^{\max}$, and the cost in USD of the maximum-scoring workflow.
 
Two patterns are noteworthy. First, on puzzles where workflows substantially outperform the strongest single-backbone baseline, for example, NaturalLanguageNavigation ($0.10$ baseline, $0.89$ workflow mean) and Sudoku2 ($0.98$ baseline, $0.95$ workflow mean), the workflow-level maximum approaches or matches the empirical ceiling, suggesting that the \aflow{} search saturates these puzzles.
Second, on puzzles where all three baselines score zero (Slant, LightUp, Minesweeper, EightPuzzle, NinePuzzle, FifteenPuzzle, SixteenPuzzle), workflows reach at most $0.05$ on Slant and LightUp, while the four sliding-tile puzzles remain at $0.00$ throughout and are excluded from the analysis corpus.

\section{Computational Cost Analysis}
\label{apdx:cost}
 
Table~\ref{tab:cost_per_instance} reports the average cost in USD to evaluate one workflow against a single puzzle instance, broken down by puzzle and difficulty. Difficulty has a small but consistent effect on cost: the cross-puzzle mean rises from $\$0.0061$ (Easy) to $\$0.0067$ (Hard). The 20-instance validation bundle (one round), therefore, costs $20\times$ the per-instance figure. Assuming the average cost per instance, the evaluation of the workflows from one \aflow{} search is in the order $\$0.0064 \times 20 \times 20$, as such for $1,740$ searches, one can expect $\sim\$4,454$, to search or evaluate the workflows. The workflow evaluation does not call the optimizer LLM, which incurs an additional cost for the search.

\section{Trace Examples}
\label{apdx:traces}

\begin{figure*}[t]
    \centering
    \includegraphics[width=\linewidth]{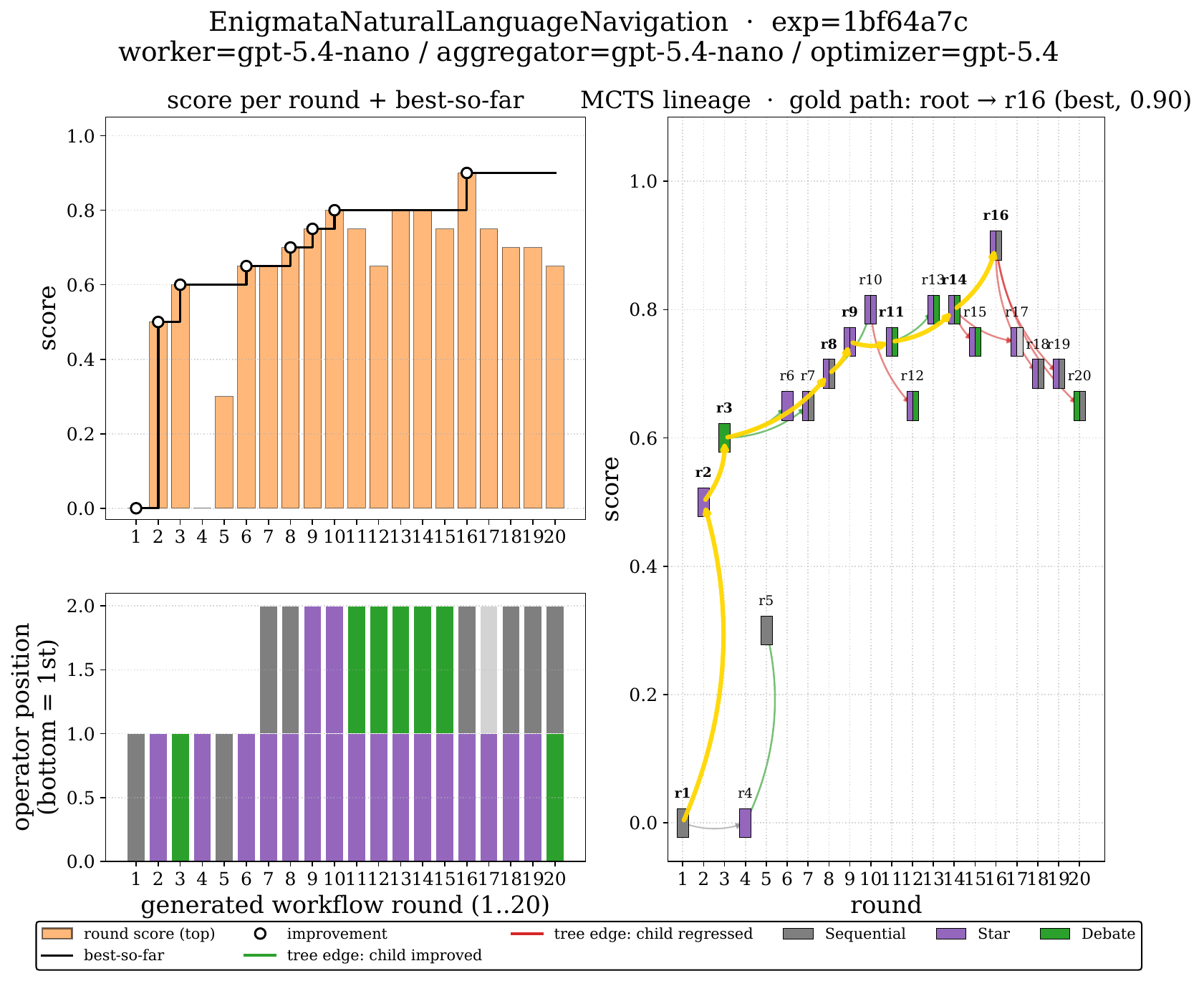}
    \caption{\aflow{} search trace for a single $(\text{puzzle}, \text{model triple})$ cell
    on EnigmataNaturalLanguageNavigation, with worker and aggregator
    at \texttt{gpt-5.4-nano} and optimizer at \texttt{gpt-5.4}
    (\texttt{experiment\_id} = \texttt{1bf64a7c}).
    \textbf{Top-left:} per-round workflow score (orange bars) and best-so-far
    envelope (black step line); open circles mark rounds that improved the
    running maximum.
    \textbf{Bottom-left:} operator-composition trajectory, with bar height
    equal to the number of operator instantiations and color denoting
    operator type (gray = \textsc{Sequential}, purple = \textsc{Star},
    green = \textsc{Debate}).
    \textbf{Right:} MCTS lineage tree, with each node labeled by round
    and positioned vertically by score. The gold path traces the
    root-to-best lineage ($\text{root} \to r_{16}$, score $0.90$); edge
    color encodes whether the child improved (green) or regressed (red)
    relative to its parent.}
    \label{fig:single_aflow_trace}
\end{figure*}

The search trace shown in Fig.~\ref{fig:single_aflow_trace} corresponds to a single
search on a NaturalLanguageNavigation bundle
puzzle on which workflow-level performance substantially exceeds the
single-backbone baseline (Table~\ref{tab:baselines}: baseline
$\le 0.15$, workflow mean $0.89$). 
 
\paragraph{Per-round and best-so-far divergence}
The per-round score and the best-so-far envelope diverge substantially throughout the search: the running maximum reaches $0.90$ at round $16$, yet four of the final five rounds score at or below $0.75$. 
This off-envelope variance is one motivation for our full-trace sampling protocol: retaining all $N = 20$ workflows from each search rather than only the argmax preserves the design-space signal carried by the intermediate and post-peak rounds, which an argmax-only corpus would discard.
 
\paragraph{Operator-composition trajectory}
The bottom-left panel shows a transition from single-operator workflows in
rounds $1$--$6$ to two-operator compositions thereafter.
During the single-operator phase, the optimizer cycles through \textsc{Sequential},
\textsc{Star}, and \textsc{Debate} as the base topology; from round~$7$ onward, the search settles into two-operator chains and explores variants within this richer topology class for the remainder of the trace.
 
\paragraph{MCTS lineage and within-search dependence}
The right panel makes the dependence structure of the search explicit. The gold path links the root to several intermediate improvements
to the peak at $r_{16}$, with each step on the path corresponding to a strict score improvement (green edges); off-path explorations such as
$r_{4}$, $r_{5}$, $r_{12}$, and $r_{17}$ either regress relative to their parent or fail to advance the running maximum.

\section{Gate Calibration}
\label{app:gate-calibration}

\begin{figure*}[!t]
    \centering
    \includegraphics[width=\linewidth]{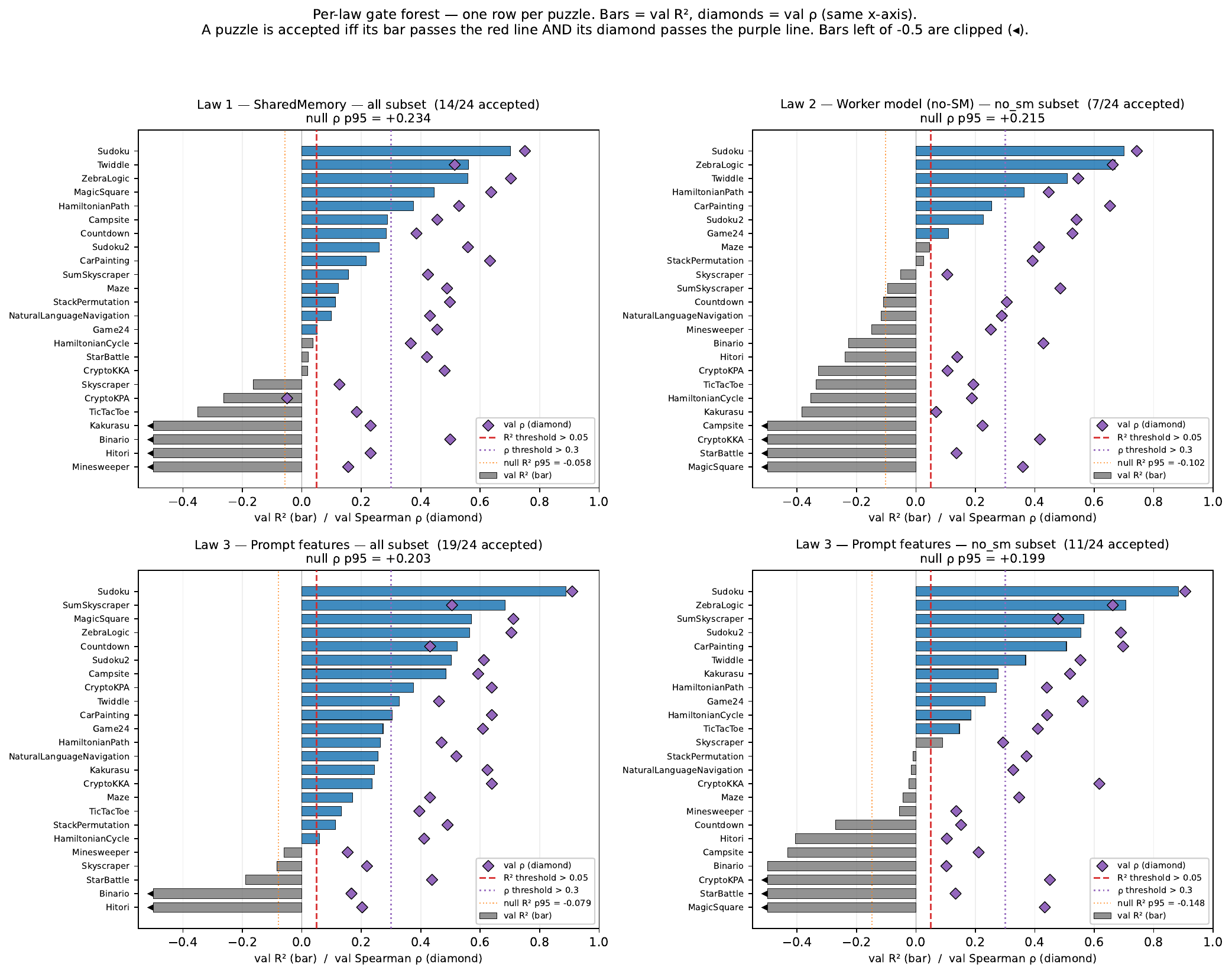}
    \caption{\textbf{Per-puzzle predictability gate diagnostics across feature subsets.} For each puzzle, bars show validation $R^2$ and diamonds show validation Spearman $\rho$ against the permutation-calibrated thresholds ($R^2 > 0.05$ red dashed line; $\rho > 0.30$ purple dotted line; 95th-percentile permutation null bounds shown in gray). A puzzle is admitted iff both metrics clear their respective thresholds on $D_{\text{val}}$.}
    \label{fig:gate_forest}
\end{figure*}

The validation thresholds $R^2 > 0.05$ and Spearman $\rho > 0.30$ used in the predictability gate (\S~\ref{subsec:analysis}) were chosen to lie a non-trivial margin above the performance attainable by chance under the same fitting procedure.
To estimate this chance baseline, we construct a permutation null distribution: for each puzzle, the training-split score column was shuffled ten times, the LightGBM regressor was refit on each permuted training set, and the resulting model was evaluated on the unshuffled validation rows.
This procedure preserves the marginal distribution of design features and the validation targets, while severing the feature--score association that the regressor is intended to capture; the resulting $(R^2, \rho)$ pairs therefore characterize the signal that the fitting procedure attributes to noise alone.

The 95th percentile of the pooled null distribution, computed across all puzzles and across every feature-set variant considered during the \autoresearch{} iteration loop, satisfies $R^2_{p95} \leq -0.06$ and $\rho_{p95} \leq 0.23$ (visualized per puzzle in Fig.~\ref{fig:gate_forest}).
The admission thresholds therefore exceed the worst-case null upper tail by at least $0.07$ on $\rho$ and at least $0.11$ on $R^2$, so any admitted puzzle clears chance by a non-trivial and feature-set-invariant margin.
The choice is conservative in the sense that it prioritizes false-negative robustness, admitting fewer puzzles in exchange for stronger guarantees on those that are admitted, which is appropriate for a downstream aggregation step in which a single floor or zero-variance puzzle would otherwise dilute the across-puzzle contrast.

\section{Broader Societal Impact} \label{sec:societal_impact}

The contribution of this work is methodological rather than capability-oriented: our framework does not introduce a new agent or extend the capability frontier of any deployed system, but provides a methodology under which the determinants of agentic performance can be examined statistically. The design principles surfaced by our analysis, although not directly deployable, contribute incrementally to the general advance of agentic capability and inherit the dual-use risk common to capability research.

\section{LLM Usage} \label{sec:llm_usage}

LLM usage is a central part of this work, as it relies on automated workflow generation, which includes code generation. Furthermore, the \autoresearch{} also conducts iterative data analysis.

We additionally use LLMs to polish the writing and check typos. LLMs were also used for programming aid, providing initial, skeleton implementations that were subsequently refined by the authors.

\end{document}